\documentclass[preprint2]{emulateapj}
\usepackage{color}
\usepackage{amsmath}
\usepackage{graphicx,epsfig}
\usepackage{longtable}
\usepackage{natbib}
\usepackage{times}
\usepackage{multirow}
\usepackage{appendix}
\usepackage{longtable,pdflscape}
\usepackage{rotating}
\usepackage{url}

\DeclareGraphicsExtensions{.jpg,.pdf,.png,.eps,.ps}

\newcommand{\degs}{deg$^2$}
\newcommand{\lstar}{L$_{\star}$}
\newcommand{\mstar}{m$_{\star}$}

\def \bcssize {$\sim$80 \degs}
\def \redshiftlimit {0.75}
\def \numberclusters{764}
\def \numberclustersfive{415}
\def \numberclusterstwenty{349}
\def \medianredshift{0.52}
\def \medianrichness{16.4}
\def \newpercent{$>$85}
\def \webaddress{ \url{http://data.rcc.uchicago.edu/dataset/blanco-cosmology-survey}}
\newcommand{\classstar}{{\sc class\_star}}

\newcommand{\msun}{\ensuremath{M_\odot}}

\newcommand{\mtwohundred}{\ensuremath{M_{200}}}
\newcommand{\mfivehundred}{\ensuremath{M_{500}}}

\def\oz#1{{}}

\def\ell{l}

\newcommand{\comment}[1]{{}}

\def\Argonne{1}
\def\KICPChicago{2}
\def\CfA{3}
\def\Miss{4}
\def\KIPAC{5}
\def\SLAC{6}
\def\AAUChicago{7}
\def\STScI{8}
\def\Stanford{9}

\begin{document}

\title{The Blanco Cosmology Survey: An Optically-Selected Galaxy Cluster Catalog and A Public Release of Optical Data Products}

\slugcomment{Submitted to \apjs}

\author{
 L.~E.~Bleem,\altaffilmark{\Argonne,\KICPChicago}
 B.~Stalder,\altaffilmark{\CfA}
 M.~Brodwin,\altaffilmark{\Miss}
 M.~T.~Busha,\altaffilmark{\KIPAC,\SLAC}
  M.~D.~Gladders,\altaffilmark{\KICPChicago,\AAUChicago}
 F.~W.~High,
 A.~Rest,\altaffilmark{\STScI}
 R.~H.~Wechsler,\altaffilmark{\KIPAC,\Stanford,\SLAC}
}

\altaffiltext{\Argonne}{Argonne National Laboratory, 9700 S. Cass Avenue, Argonne, IL, USA 60439}

\altaffiltext{\KICPChicago}{Kavli Institute for Cosmological Physics, University of Chicago, 5640 South Ellis Avenue, Chicago, IL, USA 60637}

\altaffiltext{\CfA}{Harvard-Smithsonian Center for Astrophysics, 60 Garden Street, Cambridge, MA, USA 02138}

\altaffiltext{\Miss}{Department of Physics and Astronomy, University of Missouri, 5110 Rockhill Road, Kansas City, MO 64110}

\altaffiltext{\KIPAC}{Kavli Institute for Particle Astrophysics and Cosmology 452 Lomita Mall, Stanford University, Stanford, CA, 94305}

\altaffiltext{\SLAC}{SLAC National Accelerator Laboratory, 2575 Sand Hill Rd., MS 29, Menlo Park, CA, 94025}

\altaffiltext{\AAUChicago}{Department of Astronomy and Astrophysics,
University of Chicago,
5640 South Ellis Avenue, Chicago, IL, USA 60637}

\altaffiltext{\STScI}{Space Telescope Science Institute, 3700 San Martin Dr., Baltimore, MD 21218}

\altaffiltext{\Stanford}{Department of Physics, Stanford University, Stanford, CA, 94305}

\begin{abstract}

  The Blanco Cosmology Survey is 4-band (\emph{griz}) optical-imaging
  survey that covers \bcssize \ of the southern sky. The survey consists
  of two fields roughly centered at (RA,DEC) = (23h,-55d) and
  (5h30m,-53d) with imaging designed to reach depths sufficient for
  the detection of $L_{\star}$ galaxies out to a redshift of one. In
  this paper we describe the reduction of the survey data, the
  creation of calibrated source catalogs and a new method for the separation of stars and galaxies.
   We search these catalogs for
  galaxy clusters at $z \le $\redshiftlimit \ by identifying spatial
  over-densities of red-sequence galaxies. We report the coordinates,
  redshift, and optical richness, $\lambda$, for \numberclusters \
  detected galaxy clusters at z $\le$ \redshiftlimit. This sample,
  \newpercent\% of which are new discoveries, has a median redshift of
  \medianredshift \ and median richness
  $\lambda(0.4L_{\star}$) of \medianrichness.   
  Accompanying this paper we also
  release data products including the reduced images and calibrated
  source catalogs. These products are available at
\webaddress.	

\end{abstract}
 
\keywords{galaxies: clusters: general---surveys---techniques: photometric}

\section{Introduction}
\setcounter{footnote}{0}

Multi-band optical surveys provide powerful data sets with which to
study cosmology. Technological improvements in CCD imagers over the
past two decades have led to a rapid increase in the number of such
surveys. Notable examples include the large-area Sloan Digital Sky
Survey \citep[SDSS][]{york2000} and deeper, moderate-area surveys
including the Red Sequence Cluster Survey \citep{gladders05}, the
Canada-France-Hawaii Telescope Legacy
Survey\footnote{http://www.cfht.hawaii.edu/Science/CFHLS/}, and the
NOAO Deep Wide-Field Survey \citep{jannuzi99}.  A new generation of
both large ($\gtrsim 1000$ \degs) and deep surveys including the
Panoramic Survey Telescope \& Rapid Response System 
\citep{kaiser10}, RCS2 \citep{gilbank11}, the Dark Energy
Survey\footnote{www.darkenergysurvey.org} and the Subaru Hyper
Suprime-Cam project \citep{takada10} highlight further advances in the
field.

Such surveys are particularly useful for constraining cosmology with
galaxy clusters as large areas are required to obtain statistically
useful numbers of these rare, massive objects. The selection of galaxy
clusters from optical surveys has a rich history beginning with Abell's
visual identification of over-densities of galaxies in the Palomar Sky
Survey \citep{abell58} and continuing with modern techniques that identify
concentrations of the red, passively-evolving E/S0 galaxies that form
the red-sequence in clusters
(e.g., \citealt{gladders00, koester07a, hao10}; see \citet{allen11} for
a recent review).

The critical challenge for cosmology with clusters is connecting
observable cluster properties to the mass of the system. While optical selection
does provide a mass proxy --- usually in the form of a ``richness''
parameter related to the number of galaxies in the cluster --- complimentary
mm-wave or X-ray data sets 
can greatly improve cluster mass-calibration and
provide a cross-check by testing the self-consistency of the
respective observable-to-mass scaling relations
\citep{rozo12c}. Indeed, joint optical and mm-wave analyses of the
maxBCG cluster sample \citep{koester07a} with data from the
\emph{Planck} satellite \citep{planck11-5.2c} and the Atacama
Cosmology Telescope  \citep[ACT][]{sehgal12} have already revealed
tensions between the mm-wave mass and optical richness-mass
scaling relations.

The Blanco Cosmology Survey \citep[BCS][]{desai12}, an \bcssize \
multi-band optical survey, while of moderate depth, is
particularly interesting owing to the wealth of multi-wavelength data
in the survey region. In addition to being contained within the
footprint of both the South Pole Telescope \citep[SPT][]{carlstrom11} and ACT mm-wave
surveys, one of the two BCS fields contains the XMM-BCS survey 
\citep{suhada12}, a subset of a new \emph{XMM} survey, the 
XXL\footnote{http://irfu.cea.fr/xxl/}, and is itself contained within
the 100 \degs \ SPT Deep Field which features near-infrared imaging
from \emph{Spitzer} \citep{ashby13} and has been imaged with 
\emph{Herschel}'s SPIRE camera at 250, 350 and 500 microns \citep{holder13}.

In this paper we present calibrated optical source catalogs from the
BCS.  We identify galaxy clusters in these data using a
cluster-detection technique based upon the methodology of
\citet{gladders00}, which uses a matched filter in position, color and
magnitude space to identify spatial over-densities of red-sequence
galaxies. The paper is organized as follows: in \S
\ref{sec:reduction}, we provide a brief overview of the BCS and
describe our image reduction pipeline and photometric calibration. We
describe the galaxy cluster detection algorithm in \S
\ref{sec:cluster}, and measurement of cluster properties in \S
\ref{sec:remeasure}. In \S \ref{sec:tests} we discuss tests of our
algorithm on simulated catalogs and in \S \ref{sec:compare} compare
our results to previous cluster catalogs that overlap the BCS
region. Finally, we conclude in \S \ref{sec:discussion}. Where
applicable we assume a flat $\Lambda$CDM cosmology with
$\Omega_{M}=0.27$ and $h=0.71$. Unless otherwise specified, all masses
are reported in terms of \mtwohundred, where \mtwohundred \ is defined
as the mass contained within a radius $r_{200}$ at which the average density is 
200 times the critical density, and all magnitudes are reported in the 
AB system.

\section{Data Overview and Reduction}
\label{sec:reduction}
In this section we provide a brief overview of the Blanco Cosmology
Survey and describe our production of calibrated source catalogs
from the publicly available imaging data. For a more detailed
description of the survey goals and observational strategy readers are
referred to previous publications 
\citep{menanteau09,menanteau10,desai12}.

\subsection{Survey Overview}

The BCS, an NOAO large survey program (2005B-0043), is a 4-band
(\emph{griz}) optical-imaging survey that covers \bcssize \ of the
southern sky. The survey was designed to reach depths sufficient to
detect $L_{\star}$ galaxies out to a redshift of $z=1$
\citep{desai12}. The survey consists of two fields roughly centered at
(RA,DEC) = (23h,-55d) and (5h30m,-53d), and the field locations were
coordinated to overlap with the initial mm-wave survey fields of the
South Pole Telescope and Atacama Cosmology Telescope.

The data presented here were acquired during 57 nights split over 6
observing runs between November 2005 and November 2008 using the
MOSAIC II imager\footnote{http://www.ctio.noao.edu/mosaic/} on the
4-m Blanco Telescope at Cerro Tololo Inter-American Observatory,
Chile. The MOSAIC II camera is composed of 8 2k $\times$ 4k SITe CCDs with a
plate scale of 0.27\arcsec \ per pixel, resulting in a 36\arcmin \
$\times$ 36\arcmin \ field of view. The camera was operated in 16
channel mode  (in which each CCD is read out with 2 amplifiers)
throughout the survey.

\begin{figure*}
\epsscale{1.0}
\plotone{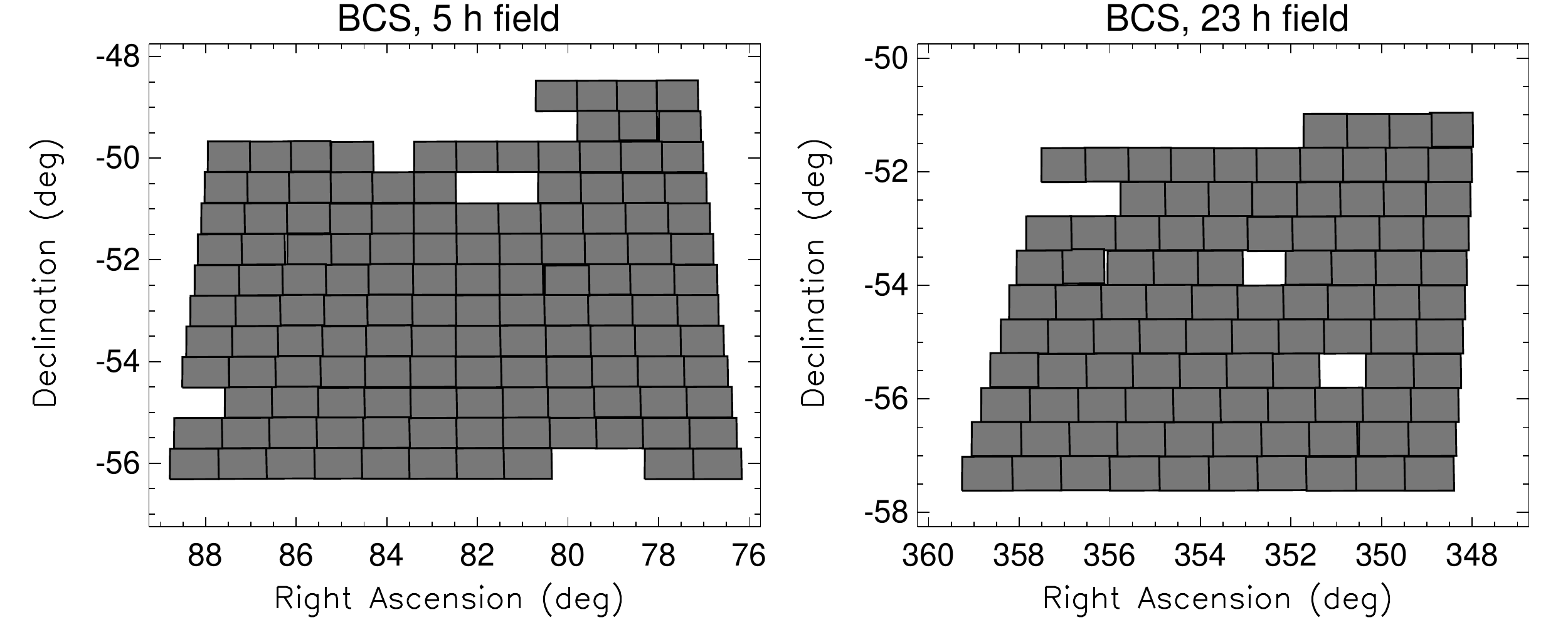}
\caption{
Spatial footprint of the imaging data presented in this work.  Coadded
images and inverse-noise weight maps for each tile are available in the
\emph{g-,r-,i-} and \emph{z-}bands. 
}
\label{fig:footprint}
\end{figure*}

The two large survey fields were observed in small tiles roughly the
size of the MOSAIC II field of view.  Here we present reductions of
133 tiles for the 5 h field and 100 tiles for the 23 h field covering
$\sim45$ \degs and $\sim33$ \degs, respectively.\footnote{We have
adopted the original field names in the NOAO archive for our tile
naming scheme and note that our names can differ from \citet{desai12} 
for the same region of sky.}. The footprint of the reduced tiles is
shown in Figure \ref {fig:footprint}.  Each individual tile nominally
consists of 2 $\times$ 125 s, 2 $\times$ 300 s, 3 $\times$ 450 s and 3
$\times$ 235 s exposures in the \emph{g-,r-,i-} and \emph{z-}bands
respectively, though the actual number of exposures can vary owing to
variable observing conditions over the course of the survey.  
Exposures are offset several arc minutes to cover chip gaps and to 
provide overlap between neighboring tiles.

\subsection{Data Reduction}

The imaging data is reduced with the PHOTPIPE pipeline. This
pipeline, initially developed for the SuperMACHO and ESSENCE projects
is described in detail in \citet{rest05a} and \citet{miknaitis07}. For all images
the reduction process includes masking of bad and saturated pixels,
crosstalk correction, overscan correction, debiasing, flat fielding,
and illumination corrections. The \emph{i}- and \emph{z}-band images
are also corrected for fringing. Illumination corrections are
determined on a nightly basis by combining all science exposures from
a given filter into a master flat. Corrections are applied from
nearby nights for nights with insufficient exposures (we find a
minimum of 11 images without bright stars is necessary to a create
good master flat). Fringe patterns on images from the MOSAIC II camera
are quite stable and we obtain good results defringing the \emph{i-}
and \emph{z-}band images using fringe frames constructed from all
science exposures obtained in the respective filter during an
observing run.

\begin{figure}
\epsscale{1.0}
\plotone{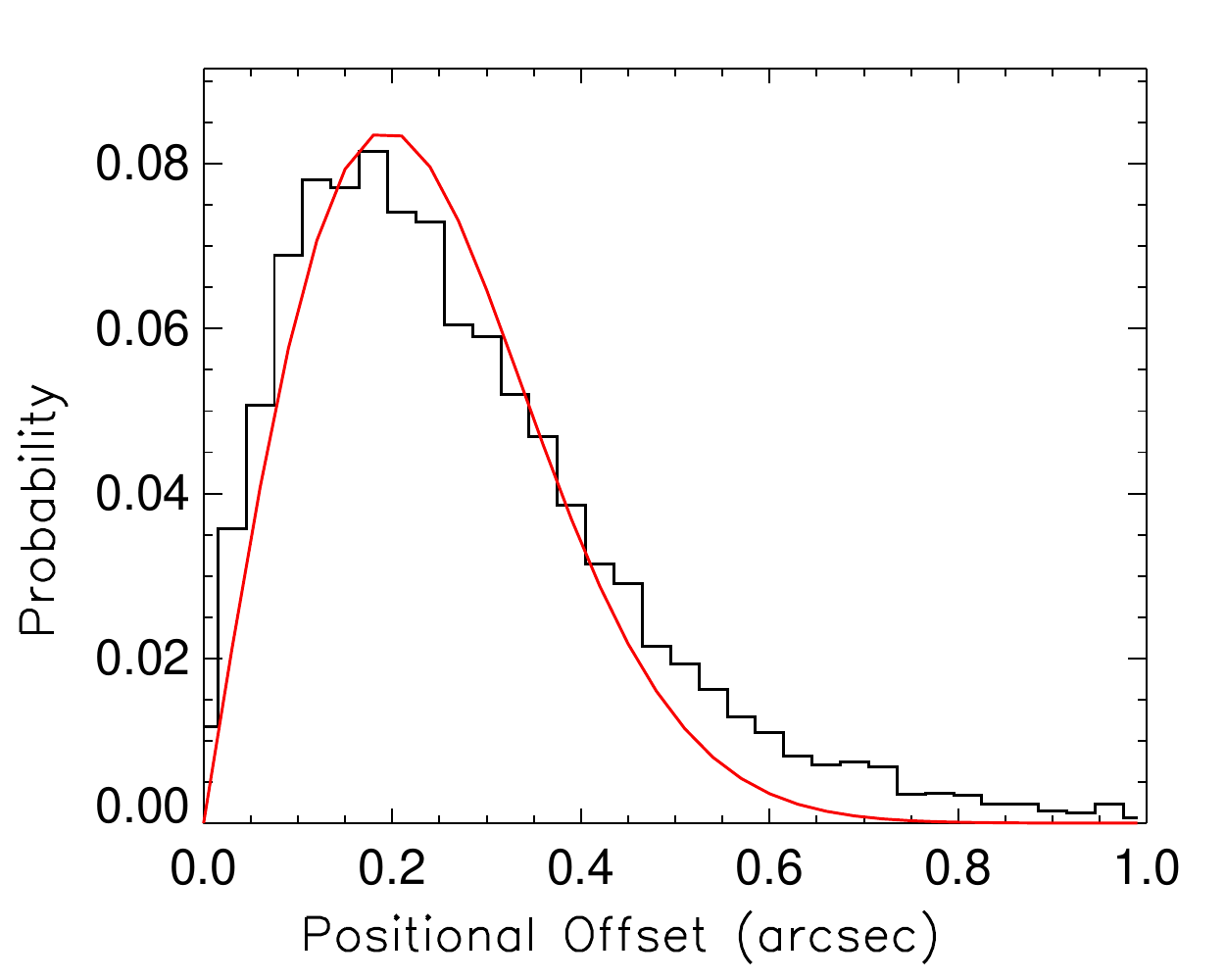}
\caption{ Astrometric residuals of BCS sources associated with 2MASS
 sources. Plotted are the residuals of sources with 16 $<$ \emph{J}$<$
 16.5 (Vega) and \emph{J}-band uncertainty $\le$ 0.25.  Overplotted is
 the best fit Gaussian model with 194 mas positional uncertainty. We
 restrict the magnitude range of the 2MASS data plotted for clarity
 as the positional scatter is a strong function of magnitude
 (signal-to-noise) at the faint end of the 2MASS catalog.}
\label{fig:astrometry}
\end{figure}

Following this processing an initial source finding run is performed
using {\tt SExtractor} \citep{bertin96}. Astrometric calibration is
then tied to the Two Micron All Sky Survey (2MASS) catalog
\citep{skrutskie06}. The astrometric solution shows residuals of $\sim$200
milliarcseconds (mas),  see Figure~\ref{fig:astrometry}. 
This is roughly the same size as the  positional uncertainty of
the typical 16.3 (Vega) \emph{J-}band 2MASS sources that are associated
with the BCS sources.

\begin{figure*}
\epsscale{1.0}
\plotone{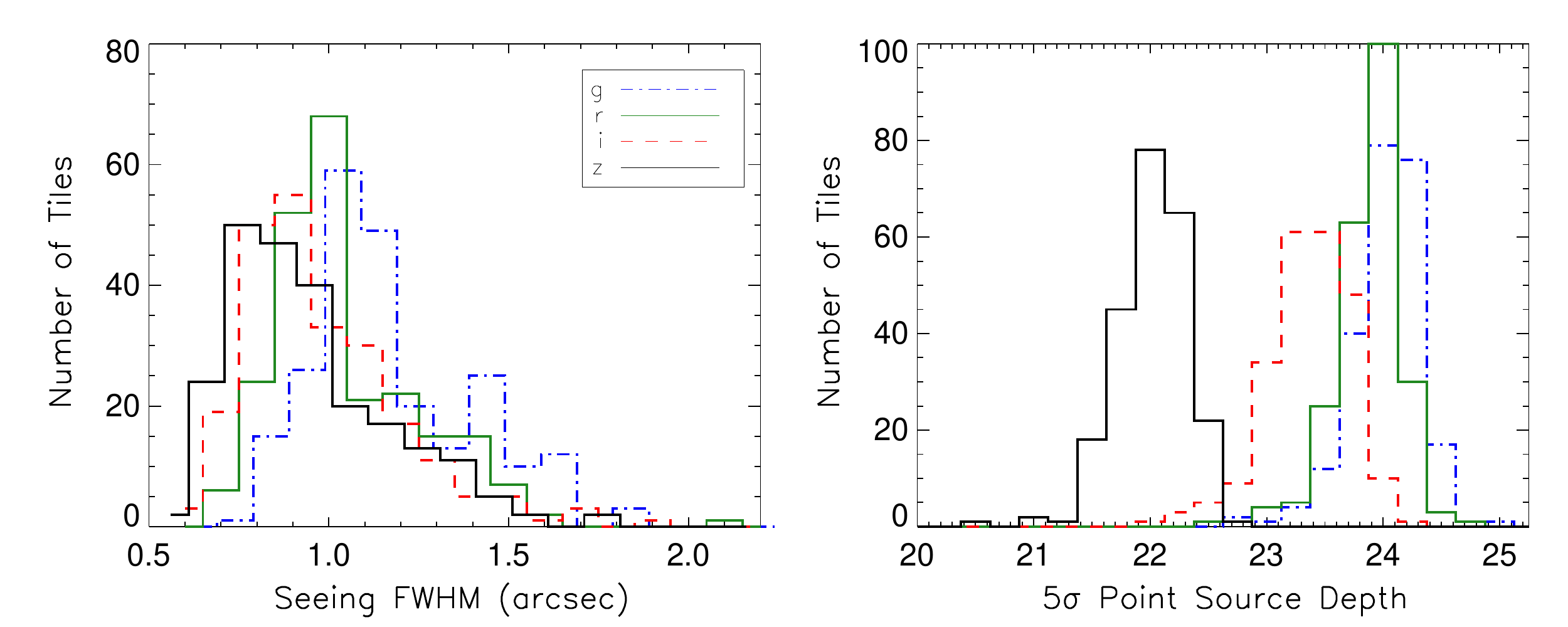}
\caption{ Seeing distribution and 5$\sigma$ corrected MAG\_AUTO point
  source depths for the 233 tiles presented in this work.  The
  plotting scheme is the same in both panels: \emph{g}-band is traced
  by blue dot-dash, \emph{r}-band is solid green, \emph{i}-band is
  dashed red and \emph{z-}band is solid black.  In the left panel the
  \emph{g-} and \emph{z-}band distributions are offset slightly to the
  right and left, respectively, for clarity.}
\label{fig:bcs_seeing_depth}
\end{figure*}

Next, in preparation for coadding the images, the relative zeropoints
of all images for each filter for a given tile are determined using
high significance objects from the initial source finding run. The
individual images are then reprojected to a common center per tile with
a pixel scale of 0.3\arcsec \ per pixel and coadded using {\tt
  SWarp}~\citep{bertin02}. 
Inverse noise-variance maps used as weight maps during final source
detection are also generated at this stage.  The distribution of
seeing ---computed as the average full width at half maximum of
the seeing disk in the single epoch images that compose the coadd---
for each of the BCS tiles as well as the 5 $\sigma$ point source
depths for the coadded images are shown in Figure
\ref{fig:bcs_seeing_depth}.

After coaddition, the tiles are visually inspected and areas with
significant artifacts (predominately corresponding to halos around the
very brightest stars or noisy amplifiers) are identified and weight
maps in these areas are set to zero. As the centers of tiles in the
different filters are sometimes slightly offset, the coadded images
are then slightly trimmed to 36\arcmin \ $\times$ 36\arcmin \ to help
ensure sky coverage in all 4 filters.  For $\sim1$ \degs \ of survey
data (distributed over many tiles) we have excluded \emph{g-}band data
owing to excess noise on the CCDs.  Finally, as discussed below, we
create flag images for use in the source extraction step based on the
weight maps, flagging regions with weight less than 10$\%$ of the
maximum weight.

\subsection{Source Catalog Creation} 
\label{sec:catalog}

Source catalogs are created by running {\tt SExtractor} 2.8.6 in dual
image mode using the \emph{i-}band images for detection and extracting
\emph{griz} MAG\_AUTO and 4\arcsec \ aperture magnitudes at the
locations of detected sources. Detection settings are provided in Table
\ref{tab:sextractor_settings}.

\begin{deluxetable}{ll}  
\tablecolumns{2}
\tablecaption{SExtractor Source Detection Settings
\label{tab:sextractor_settings}
} \tablehead{ 
} \startdata DETECT\_TYPE & CCD \\ DETECT\_MINAREA & 1.1$\pi\times$
(\emph{i}-band seeing)$^{2}$\\ THRESH\_TYPE & RELATIVE \\ DETECT\_THRESH
& 1.2 \\ ANALYSIS\_THRESH & 1.2 \\ FILTER & Y \\ FILTER\_NAME &
default.conv \\ DEBLEND\_NTHRESH & 32 \\ DEBLEND\_MINCONT & 0.005 \\
BACKPHOTO\_TYPE & LOCAL \enddata \tablecomments{SExtractor settings
used in the extraction of source catalogs. SExtractor was run in dual
image mode with the \emph{i}-band images set as the detection
images. The \emph{i-}band seeing value refers to the average full
width at half maximum of the seeing disk in the single epoch images
that compose the coadds.}
\end{deluxetable}

Following the creation of the source catalog we perform additional
steps before photometric calibration. First, we identify regions of
roughly uniform coverage in each coadd. As described above, the BCS
tiles are nominally composed of 2 or 3 exposures in each filter, with
offsets of several arc minutes between exposures. As we are processing
the survey on a tile-by-tile basis (as opposed to coadding all images
of a given filter into a single monolithic block for each survey
field) this naturally leads to shallower regions at the edges of the
tiles. Based on the source location in the tile we designate each
source as coming from the central (i.e.  roughly uniform coverage) or
edge region of the tile.

We next apply a correction to the magnitude uncertainties as
we find that the default SExtractor uncertainties underestimate the
scatter in the data. We estimate a correction factor by measuring the
sky noise ---the dominant contribution to the flux uncertainty for
faint sources--- utilizing a modified version of the Monte-Carlo based
technique described in \citet{ashby09}. In brief, we perform
photometry on the sky in 4\arcsec \ apertures at 1500 random positions
in the central region of each coadded image. We then fit a Gaussian
function to the resulting flux distribution, excluding the bright tail
which is biased by real sources in the image. We compare the measured
scatter to the median 4\arcsec \ flux uncertainty estimated by
SExtractor for sources from this same region. We rescale the 
error bars by this factor which is typically 1.3 to 1.5 times greater
the initial uncertainty estimated by SExtractor.  We have verified
this correction by comparing the distribution of magnitude differences
for objects measured in sequential single epoch imaging with the
corrected magnitude uncertainties. Our rescaling factor also agrees
with the correction determined in \citet{brown07} who found a 40\%
underestimate in the MAG\_AUTO magnitude uncertainties.  
This method works for sources with signal-to-noise $\gtrsim 2.5$, below 
which there is again extra scatter not reflected by the rescaled error
bars. The 5 $\sigma$ point source depths plotted in Figure
\ref{fig:bcs_seeing_depth} are derived from corrected uncertainties on
point sources.  We find for the detection band, \emph{i}-band, these
depths correspond to the peaks in number counts distributions.

The last step we perform before photometric calibration is identifying
sources that may have biased flux measurements owing to their
proximity to bright stars. Such stars are often surrounded by
``halo-like'' features in the image caused by internal reflections in
the camera. We find it necessary to flag regions around stars brighter
than 14 (Vega) in \emph{J-}band. At fainter magnitudes default
flagging from the weight maps is sufficient. All stars with
\emph{J-}band magnitude brighter than 7 (Vega) are visually inspected
in pseudo-\emph{rgb} color images generated from the
\emph{z-,i-,r-}band coadded images and the affected regions are
flagged. Fifty stars in the 23 h field and 101 stars in the 5 h field
meet this magnitude cut. For the fainter stars flagged regions were
automatically generated in circular apertures around each star. The
radii of these apertures were conservatively chosen based on
inspection of a subset of stars. For these fainter stars the aperture
radii ranged from 2\arcmin \ at J=7 (Vega) to 0.25\arcmin \ at J=14
(Vega). Sources in these flagged regions are excluded when fitting for
the photometric calibration parameters for each tile (described
below), but are retained in the released catalogs and marked as having
potentially biased photometry.

We calibrate the colors of stars and galaxies using Stellar Locus
Regression \citep[SLR][]{high09}. SLR calibrates colors by matching the 
instrumental colors of stars to that of a universal stellar
color-color locus as measured by $\sim10^{5}$ stars in SDSS \citep{covey07}.  
SLR has previously been
used to calibrate the photometry for pointed follow-up of SPT-detected
galaxy clusters  \citep[e.g.,][]{high10}, an alternate reduction of the
BCS \citep{desai12}, and similar techniques have been utilized
elsewhere in the literature for the photometric calibration of large
surveys \citep{gilbank11} and to compliment more standard calibration
\citep{ivezic07}. As discussed in \citet{high09}, SLR naturally
corrects for atmospheric and ---as the majority of the stars at BCS
survey depths are behind the Galactic dust sheet--- Galactic
extinction.  To determine the necessary calibration terms on a
tile-by-tile basis, for each tile we use sources with signal-to-noise
greater than 3 in the color combinations of interest, SExtractor
\classstar \ greater than 0.95 in the \emph{r-} and \emph{i-}bands and
require the star to be located in the central region of the coadded
images. By using only the central portion of the coadd to determine
the photometric calibration, we can use the overlap regions at the
edges of the tiles as an independent check of our calibration.

We transform the MOSAIC II magnitudes to the SDSS system using a
single set of color terms for all CCDs (Table \ref{tab:color}). The
color terms were measured by performing a first order fit between
MOSAIC II instrumental and SDSS magnitudes in several SDSS-observed
star fields. When correcting for extinction and color terms, it is
necessary to chose a color multiplier in the transformation from raw
to calibrated colors. 
In general, broad filter combinations (such as \emph{g-i}) are
beneficial as they provide wide leverage over the range of stellar
colors.  However, owing to the relatively shallow depth in
\emph{g-}band of the BCS catalogs with respect to high-redshift
L$_{\star}$ galaxies (and as these catalogs were constructed for
detection of red-sequence cluster galaxies), we chose the color
multiplier to always include the \emph{r-}band filter.

\begin{deluxetable}{ccc}  
\tablecolumns{3}
\tablecaption{Color Terms
\label{tab:color}
}
\tablehead{   
  \colhead{Filter} &
  \colhead{Color Term} &
  \colhead{Color Multiplier}
}
\startdata
\emph{g} & -0.1344 &\emph{g - r} \\
\emph{r} & -0.0103 & \emph{r - i} \\
\emph{i} & -0.1148 & \emph{r - i} \\
\emph{z} & 0.0028 & \emph{r - z }
\enddata
\tablecomments{Color terms and color multipliers applied in the calibration of source
catalogs from the BCS. }
\end{deluxetable}

The absolute flux scaling is also calibrated using SLR by matching
bright stars in the BCS with stars from the 2MASS point source catalog
\citep{skrutskie06}. Determining robust zeropoints proved to be the
most difficult part of the catalog calibration process, so we discuss
the process in some detail here.  Constraining zeropoints with SLR
proceeds analogously to all other color-color calibrations with the
stellar locus, only one of the colors is constructed from an
instrumental magnitude and 2MASS magnitude. The quality of the zeropoint
fit can be checked by obtaining zeropoints for each of the filters and
comparing colors constructed from these absolute magnitudes against
colors from the SLR-color calibration.  Initial calibration (without
flagging generated from the weight maps) revealed large discrepancies
between these two sets of calibrated colors.  We found that these
discrepancies were largest for tiles observed in the best conditions
and also noticed that the brighter, bluer stars systematically
deviated from the median locus relation --- both of these symptoms
pointed to saturated stars being included in the zeropoint
calibration.

We utilize the SWarp-generated weight maps to identify the non-masked
saturated sources (which we note did not exceed the {\tt SATURATE}
keyword in the images' fits headers and so were not masked earlier in
the reduction process). The algorithm that creates these weight maps
includes a contribution from the Poisson error from bright
sources\footnote{
https://www.astromatic.net/pubsvn/software/swarp/trunk/doc/swarp.pdf},
which makes it possible to identify the locations of saturated sources
in the map prior to source extraction. We find that flagging objects
whose weight is less than 10$\%$ of the maximum weight is sufficient
to exclude saturated objects. Owing to the long exposures in
\emph{i-}band (450 s) a number of tiles had too few non-saturated
2MASS sources in the \emph{i-}band selected catalogs 
to provide accurate calibration. We instead determine zeropoints for
all tiles using the \emph{grz-}band magnitudes of these 2MASS sources.
For these bright objects we report only the \emph{grz}-band MAG\_AUTO
magnitudes and associated colors in the released catalogs.

\begin{figure*}
\epsscale{1.0}
\plotone{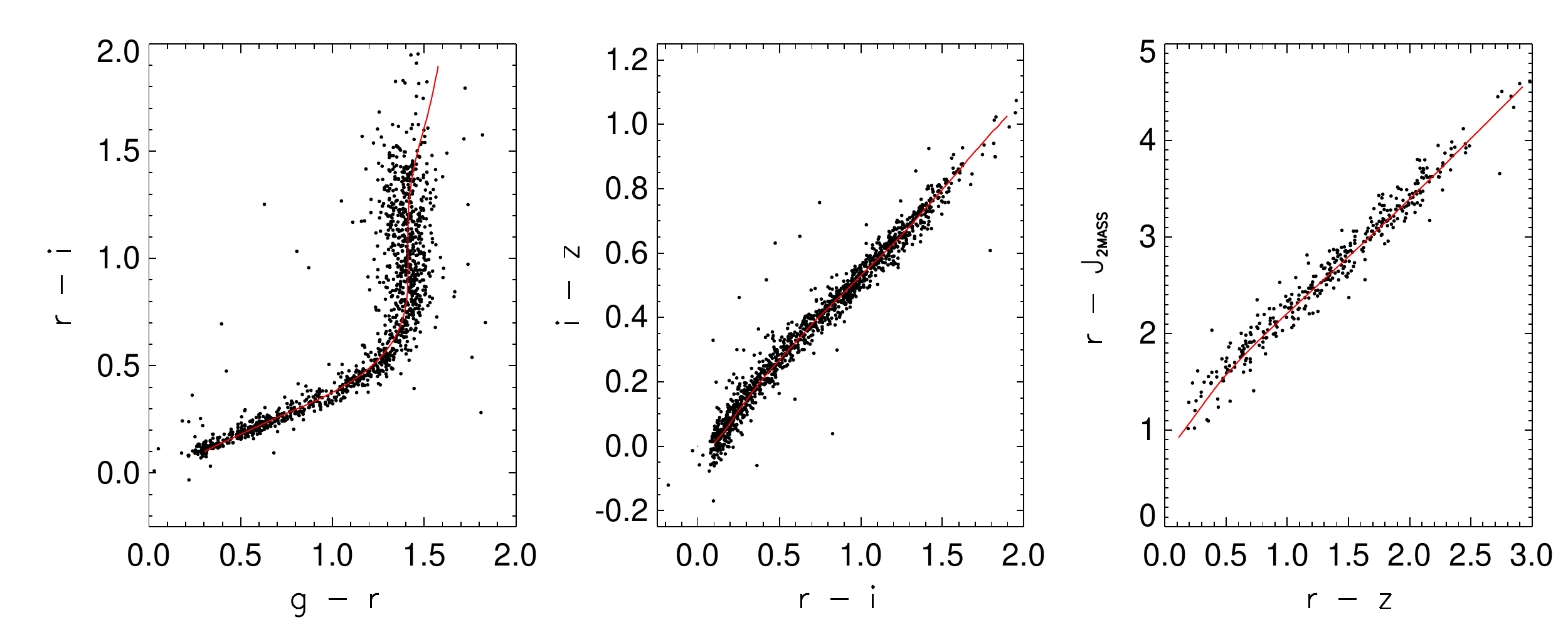}
\caption{Color-color diagram of a subsample of stars with \emph{SG}
$>$ 0.8 (see \S\ref{sec:stargal}) and color errors less than 0.25 magnitudes after photometric
calibration of BCS tile 0548-5524. Overplotted in red are the median
color-color relations of the stellar locus as reported in
\citet{covey07}.  }
\label{fig:slrplot}
\end{figure*}

Zeropoints were calculated using \emph{z-J} versus \emph{r-z} and
\emph{r-J} versus \emph{r-z}. For many tiles we found \emph{g-J}
unsuitable as, when the brightest (bluest) objects are saturated, the
locus for the remaining fainter objects is essentially vertical and
provides little constraining power. Comparison between the
\emph{i-}band magnitudes computed using SLR-derived colors and
absolute \emph{r-} and \emph{z-}band magnitudes shows good agreement
(0.02 magnitude root-mean-square (rms) in the difference).
For simplicity, we set the zeropoints of all filters using the
\emph{r-}band zeropoint calibration and the SLR-derived colors. In
Figure \ref{fig:slrplot} we show an example of the SLR calibration for
a typical BCS tile.

\subsection{Photometry Validation}
\label{sec:validation}
\begin{figure*}
\epsscale{1.0}
\plotone{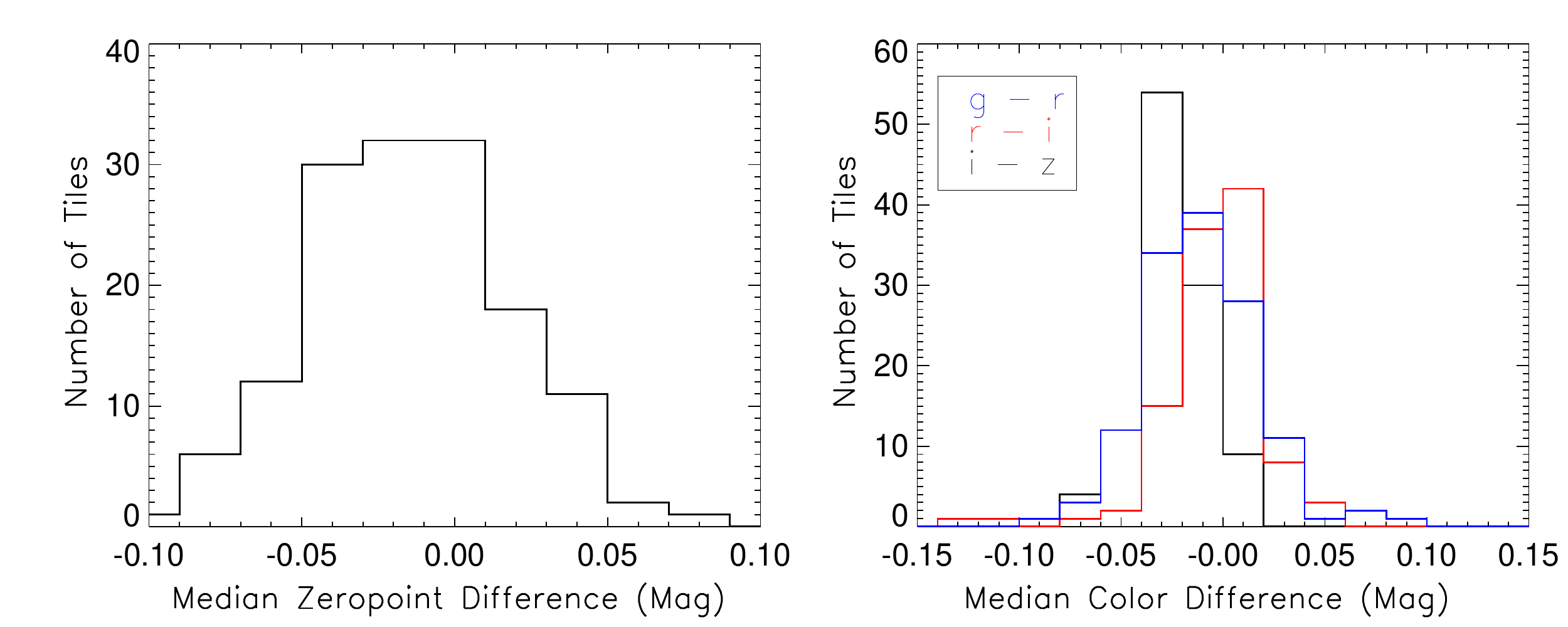}
\caption{Median \emph{i-}band magnitude (\emph{Left}) and \emph{g - r,
    r - i, i - z} color (\emph{Right}) differences for objects that
  lie in the overlapping region between neighboring tiles.  We report
  magnitude and color statistics for all tiles with $\geq 100$ sources
  with signal-to-noise $\geq4$ in the \emph{r-,i-,z-}bands ($\geq 75$
  sources with signal-to-noise $\geq4$ in \emph{g-}band) corresponding
  to 204 (188) of the 233 survey tiles. We divide the standard
  deviation measured from these distributions by $\sqrt{2}$ to obtain
  the uncertainty on the calibration for a single tile. We measure
  2.4\% rms variation in \emph{i-}band zeropoints and 1.9\%, 1.3\%,
  1.3\% rms in the \emph{g-r}, \emph{r-i} and \emph{i-z} calibration,
  respectively. }
\label{fig:photometry_calibration}
\end{figure*}

We perform several tests to check the accuracy of the catalog
calibration. First, we compare photometry for objects in the few arc
minute overlapping regions between neighboring tiles. As mentioned
above, as each tile is calibrated independently and the edge regions
are not included in the determination of the calibration solution, we
can use these regions to test the quality of our photometric
solutions. We note that, as this test is performed on the edges of the
image where corrections for flat fielding are generally the poorest,
the photometry in the central portion of the tile is potentially
better. We report statistics for the median magnitude or color
difference of objects in tiles for which this overlap region has
$\geq$100 sources with signal-to-noise $\geq$4 in the
\emph{r,i,z-}bands (204/233 tiles) or $\geq$75 sources in the
\emph{g-}band (188/233 tiles). The results are shown in Figure
\ref{fig:photometry_calibration} . We divide the standard deviation
measured from these distributions by $\sqrt{2}$ to obtain the
uncertainty on the calibration for a single tile. From these tests we
measure 2.4\% rms variation in \emph{i-}band zeropoints. Checks on the
colors show 1.9\%, 1.3\%, 1.3\% rms in the \emph{g-r}, \emph{r-i} and
\emph{i-z} calibration, respectively. Investigation of the tile pairs
with outlying median-color differences shows that the outliers
predominately correspond to tiles with poorer seeing or significantly
shallower-than-average data at the tile edges.

As another test, we use the stellar locus to check whether our
reported magnitude uncertainties accurately reflect the measurement
uncertainty. Using high signal-to-noise stars from the ``high-quality"
sample described in \citet{covey07} we identify low-scatter regions of
the stellar locus that are locally linear in various color-color
combinations. We then fit for these linear relations. For each tile in
the BCS we test the reported uncertainties using these same color
combinations.  For each color-color pair we use stars where
uncertainties in the abscissa color are small but errors in the
ordinate color are greater than the locus scatter. This requirement
ensures that the measured spread in the distribution around the
above-described linear relation is dominated by our measurement
uncertainty.  To check the \emph{i-z} color uncertainties we use stars
with \emph{g-r} \textless 1.2, for the \emph{g-r} color we use 0.5
\textless \emph{i-z} \textless 0.8 and for \emph{r-z} we use stars
with 0.2 \textless \emph{g-i} \textless 1.8.  This test proved
extremely useful ---results on preliminary catalogs uncovered a small
error in the coaddition process and identified single-epoch images to
exclude from the coadds.  The final distributions are well behaved,
with some excess scatter again observed in filters with poorer seeing
(this particularly affects the \emph{g-}band, as this band typically
had worse seeing and lower sky noise).

\subsection{Star Galaxy Separation}
\label{sec:stargal}

\begin{figure*}
\epsscale{1.0}
\plotone{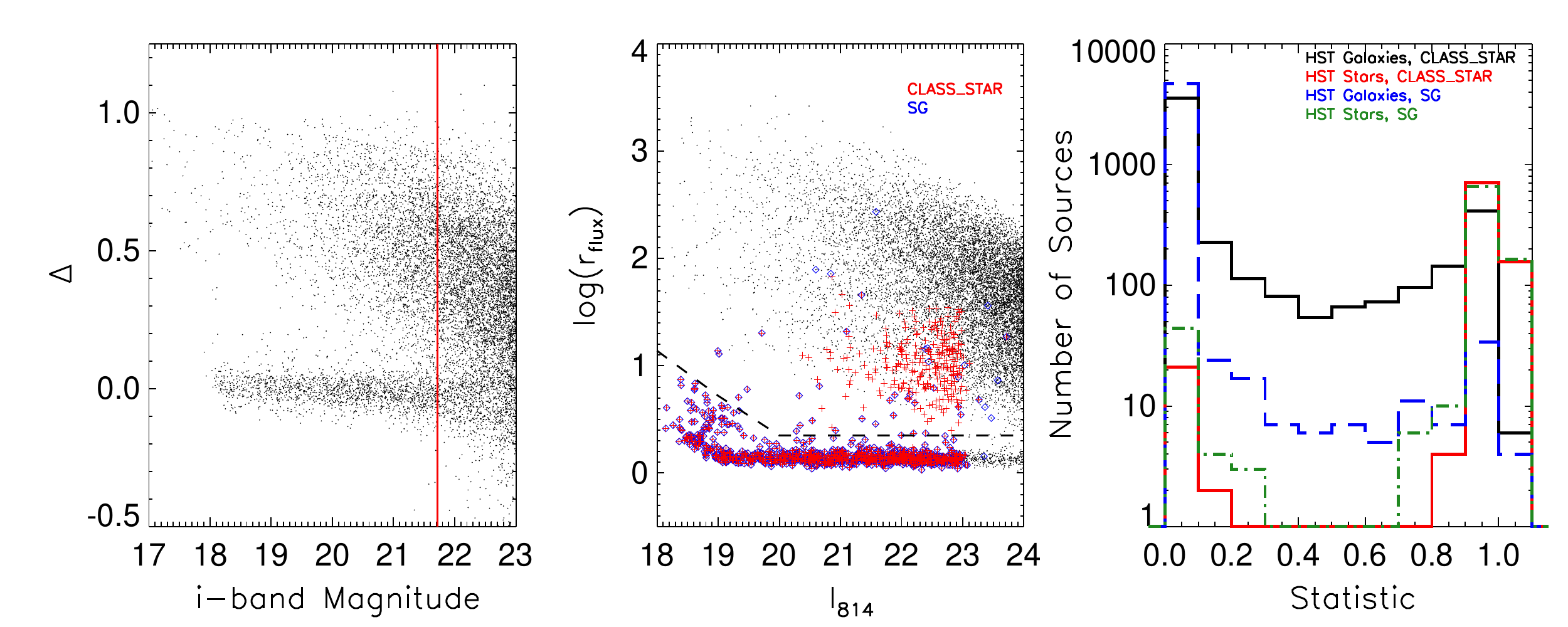}
\caption{ (\emph{Left}) Example of the new star-galaxy separation process
  discussed in \S\ref{sec:stargal} as applied to BCS tile
  0511-5448. Plotted is morphological parameter, $\Delta$, versus
  magnitude. The stellar population forms a narrow band centered
  around 0. The solid red line corresponds to the \emph{i-}band
  magnitude (21.7 for this tile) at which there is a ratio of 9 stars
  to each galaxy at the center of this band. (\emph{Middle}) Test of the new
  star-galaxy statistic using \emph{HST}-classified sources in the
  Extended Groth strip.  We classify sources below the black
  dashed-line in the I$_{814}$-log(r$_\mathrm{flux}$) plane as stars in the
  HST catalog. Overplotted in red are sources with CLASS\_STAR $\geq
  0.95$ and in blue are sources with \emph{SG} $ \geq 0.8$. The
  limiting classification magnitude (in \emph{i}-band from the
  ground-based optical data) is determined as in the previous panel.
  (\emph{Right}) Distribution of CLASS\_STAR and \emph{SG} parameters for
  \emph{HST}-classified galaxies (solid black and dashed-blue,
  respectively) and \emph{HST}-classified stars (solid red and
  dot-dashed green). The last two panels demonstrate that
  significantly fewer galaxies are erroneously removed from a galaxy
  catalog generated from a cut on the star-galaxy statistic using
  \emph{SG} compared to CLASS\_STAR.}
\label{fig:star_gal}
\end{figure*}

As a cross-check of the SExtractor star-galaxy statistic, \classstar,
and to identify the magnitude limit to which this separation is
robust, we perform an additional morphological classification of the
detected sources. The process to determine this new statistic is
simple: we square the \emph{i-}band images, source extract in dual
image mode using the original \emph{i}-band image as the detection
image and then compute the difference in magnitudes between sources in
the original and squared image. Using bright stars identified with
\classstar \ $\geq$0.98 we fit for and subtract a linear relation
between this difference and the calibrated magnitudes.  The result of
this process is shown in the left panel of Figure
\ref{fig:star_gal}.  As can be seen for bright objects this adjusted
difference, $\Delta$, between the ``squared'' and original magnitudes
separates into two populations, with $\Delta$ of stars centered around
zero.\footnote{To motivate this statistic consider the simple case
  where stars are modeled as 2-dimensional Gaussians. For a Gaussian
  distribution, the squaring and subsequent differencing procedure
  described above produces a linear relation between magnitude -
  magnitude$_{\rm squared\_image}$ versus magnitude with a slope of
  -1. Subtracting off this relation then centers the stellar
  population around zero. Galaxies, with considerably more extended
  profiles, form a separate population until noise and limited
  resolution conspire to wash out morphological information.}  We find
the separation between the two populations extends to fainter
magnitudes using aperture photometry instead of MAG\_AUTO, so we
calculate the difference using 3\arcsec \ aperture magnitudes. As there
are few stars brighter than 19.5 \emph{i-}magnitude and \classstar \
is robust at this bright end in all tiles, we use the \classstar \
statistic for these sources. For sources fainter than 19.5, we fit a
double Gaussian model to the sources in sets of 400 ordered by
increasing magnitude. The new star-galaxy statistic, \emph{SG}, is
defined as:
\begin{equation}
SG = \frac{A_\mathrm{star} \ e^{(\Delta - \mu_\mathrm{star})^{2} \over 2\sigma_\mathrm{star}^{2}}}{A_\mathrm{star} \ e^{(\Delta - \mu_\mathrm{star})^{2} \over 2\sigma_\mathrm{star}^{2}} + A_\mathrm{galaxy} \ e^{(\Delta - \mu_\mathrm{galaxy})^{2} \over 2\sigma_\mathrm{galaxy}^{2}}}
\end{equation}
where the parameters \emph{A}, $\mu$, and $\sigma$ correspond to the
amplitude, mean and standard deviation of the Gaussian fits, and ---as
with \classstar --- a value of 1 corresponds to a high probability of the
source being a star. Finally, for the few sources with $\Delta$
\textless 0 and \emph{SG} \textless 0.8 we set \emph{SG} to 2.0 as
visual inspection of these objects indicate they are predominately
image artifacts.

For each BCS tile, we record this new statistic for each tile up to
the magnitude where there is a ratio of 9 stars to each galaxy at the
mean of the Gaussian that corresponds to the stellar population. As
for any morphological classifier, the magnitude limit to which this
classifier is robust is highly seeing and depth dependent. We include
this limit as a column in Table \ref{tab:tile}.

To evaluate the performance of the new star-galaxy statistic we use
external optical- and spaced-based imaging from the Extended Groth
strip.  Using the same SExtractor settings as in the BCS, sources are
extracted from deep, ``best-seeing" (0.65\arcsec\ FWHM) \emph{i}-band
coadds from the Canada-France-Hawaii Telescope Legacy Survey
\citep{gwyn12}.  Owing to the excellent seeing in these coadds, we
compute the \emph{SG} statistic using 1.5\arcsec \ aperture magnitudes
to better demonstrate the power of this statistic as a morphological
classifier (while the 3\arcsec\ aperture-derived statistic proved
robust in this test, the limiting magnitude was roughly 0.75
magnitudes shallower as the larger aperture did not fully leverage the
available spatial resolution).  Using a 1\arcsec\ association radius,
we match this source list with the Advanced Camera for Surveys
General Catalog \citep{griffith2012} which contains morphological
information derived from \emph{HST} imaging.  Guided by
\citet{gray2009} (Equation 1), we adopt an \emph{HST}-based
morphological classification using the SExtractor MAG\_BEST and
FLUX\_RADIUS ($r_\mathrm{flux}$) parameters derived from the
$I_{814}$-band. We classify as stars sources with
\emph{$i$}-band $< 23$ (the \emph{SG}-magnitude limit determined as
described above) and
\begin{align*}
\mathrm{log}(r_\mathrm{flux}) < \mathrm{max}(0.35, 1.50-0.05(I_{814}+1), 8.7 - 0.42I_{814}) . 
\end{align*}
The resulting $\sim$900 sources matching these criterion (out of
$\sim$5,000 total sources brighter than the magnitude cut) are used as
the ``true" stars in our evaluation of the ground-based morphological
classifiers.  

We plot comparisons of the CLASS\_STAR- and \emph{SG}-classifications
to the \emph{HST}-results in the middle and right panels of Figure
\ref{fig:star_gal}.  As demonstrated in these panels, the
\emph{SG}-statistic shows significantly fewer galaxies misclassified
as stars. In this test, considering only sources brighter than the
classification magnitude (\emph{i}-band=23), a catalog generated from
\emph{SG}$<0.8$ contains $99$\% of all galaxies and removes 93\% of
all stars, while a similar catalog from CLASS\_STAR $<0.95$ includes
$94\%$ of all possible galaxies and excludes $95\%$ of all stars.

\section{Cluster Detection}
\label{sec:cluster}
We detect clusters using an algorithm based upon the red-sequence
cluster detection algorithm presented in \citet{gladders00,
gladders05}.  The algorithm identifies clusters as over-densities of
galaxies in position, color and magnitude space.  Here we provide a
brief overview of the algorithm and note key differences in our
implementation as compared to the literature.

Similar to \citet{gladders00, gladders05}, the algorithm works by
assigning each galaxy a weight as a function of redshift where weights
are based upon the consistency of the galaxies' magnitudes and colors
with a model for the red sequence as a function of redshift.  The
weighted density of galaxies in a series of redshift slices is
computed in 0.25\arcmin \ cells and smoothed with a kernel with a core
radius of 350 kpc. The galaxies that contribute non-zero weight in
each slice are then bootstrap resampled and new density maps are
created to estimate the background distribution that is used to map
the density values to Gaussian significance. The significance maps at
each redshift are then stacked into a 3 dimensional datacube in which
peaks greater than a significance threshold of 3.1$\sigma$ and with at
least five galaxies contributing any amount of weight to the detection are identified as clusters
(the additional constraint is necessary to reduce spurious detections at
high-redshift). 
This threshold, determined using mock catalogs (described in \S \ref{sec:tests}), was chosen to limit catalog
impurity (including the failure mode of selecting small clumps of
large clusters as separate systems) while maximizing completeness.

The largest change in our implementation of the cluster-finding
algorithm lies in the procedure for placing the small 0.33 \degs \
tiles into larger blocks for
the creation of the background estimates that define the detection
significance. As the BCS survey is somewhat heterogeneous (Figure
\ref{fig:bcs_seeing_depth}) , it is difficult to apply the rigorous
image matching technique of \citet{gladders05} and still obtain blocks
of sufficiently large area for robust background estimation. Instead,
we simplify the problem by restricting our cluster search to lower
redshifts and brighter magnitudes where the differences in photometric
errors between tiles are reduced. As part of this process we restrict
the magnitude weighting applied during the assignment of galaxy
weights to magnitudes brighter than m$_{\star}$ + 1.5 in each redshift
slice.

To define the regions for the cluster search we begin by excluding
tiles with \emph{i}- or \emph{r-}band seeing in excess of 1.6\arcsec.
In the 5 h field we also exclude 6 tiles (BCS0532-5412, BCS0532-5448,
BCS0536-5448, BCS0536-5412, BCS0540-5412 and BCS0540-5524) owing to
prominent Galactic cirrus visible in the tiles. We then group the
tiles based upon their noise properties in the \emph{r-, i-} and
\emph{z-}bands. For each field we produce three different catalogs --
a large catalog used in a ``low-redshift'' cluster search to \emph{z}
= 0.5 and two smaller catalogs for a higher redshift cluster search
extending to \emph{z} = 0.84 (above our nominal redshift
upper limit of \redshiftlimit ).  For each catalog we ensure roughly
uniform completeness levels at m$_{\star}$ + 1.5 (\emph{i-}band) at
the highest redshift of interest by excluding tiles that have
\emph{i-}band 5 $\sigma$ point source depths less than m$_{\star}$ +
1.75 at this redshift.

The ``low-redshift'' catalogs for each field are created by merging
source catalogs from tiles satisfying the above constraints and the
additional requirement that a typical m$_\star$ + 1.5 galaxy at z =
0.5 have a \emph{r-z} color uncertainty less than 0.16 (80/100 tiles
23 h field, 115/133 tiles in the 5 h field). We run two searches on
these catalogs -- first using the \emph{g-, r-, i-}band magnitudes we
detect clusters between redshifts 0.15 and 0.35.  We then extend the
search to redshift 0.5 using the \emph{r-, i-, z-}band magnitudes.  
The two higher redshift catalogs for each field are created by merging
the source catalogs of the tiles that have \emph{r-z} color
uncertainty of a red-sequence m$_{\star}$ + 1.5 galaxy at z = 0.7 less
than 0.33 ( 32/100 tiles 23 h, 62/133 5 h) and between 0.33 and 0.4
(31/100 tiles and 34/133 tiles in the 23 h and 5 h fields
respectively). In Table \ref{tab:tile} we flag the tiles searched for
clusters.

 As the tile boundaries can overlap at the edges, when creating the
merged catalogs we check for and remove duplicate sources. We identify
the sources for a given tile which have a counterpart within an
adjacent tile using a 1\arcsec \ association radius. When duplicates are
found, we use the photometry with the smaller \emph{r-i} color
uncertainty in the final merged catalog.

Once the tiles are chosen, we use the new star-galaxy statistic,
\emph{SG}, to separate stars and galaxies for sources brighter than
\emph{i}=20.5 (corresponding to m$_\star$ at z = 0.55), classifying
sources as stars when \emph{SG} $\ge$0.8. We choose this conservative
magnitude for separation to mitigate spatial variations in source
density induced by seeing variations in the stellar-excised source
catalogs.

We next identify regions to utilize for background estimation ---
these regions are slightly more restrictive than the actual area used
for the cluster search. Regions for inclusion are identified in a two
step process: first, using the weight maps produced by SWarp for each
tile during the coaddition process, we mark as `good' the 0.3\arcsec \
pixels that have greater than one-third of the median weight, lie
inside the nominal central region of the tile, and are not in a region
previously flagged during the cataloging process (see Section \S
\ref{sec:catalog}).  We next rebin these small pixels into the
0.25\arcmin \ -scale pixels that are used by the cluster detection
algorithm and mask the 0.25\arcmin \ pixels for which less than 75\%
of the pixel area is marked for inclusion. Density values are drawn
from unmasked pixels during the bootstrap resampling step. We include
the edge regions of the tiles in the cluster search as they provide a
natural taper at the edge of each tile but exclude them from the
bootstrapping process as the coverage at the edges of tiles is
significantly more variable than in the central regions.  As a
consequence the detectability of clusters in these edge regions is
lessened owing to both the higher photometric scatter of sources in
these regions and as they are less well-matched to the estimated
background drawn from the less noisy central regions.

Finally, the red sequence models used in this analysis were created
using the {\tt GALAXEV} routines provided in conjunction with
\citet{bruzual03}.  The models consist of a passively-evolved single
stellar population generated with the Salpeter initial mass function,
the Padova 1994 tracks and an instantaneous star burst at redshift z =
3.  Metallicites are chosen based upon analytical fits to RCS2 cluster
data (Koester, \emph{private communication}). Cubic splines are used
to interpolate the discrete output of the code to arbitrary redshifts.
Assuming passive evolution, we tie our $m_{\star}(z)$ model to the
analytical function valid $0.05 < z < 0.35$ for the maxBCG cluster
sample as reported in equation 11 in \citet{rykoff12}.

We calibrate three color-magnitude relations as a function of
redshift: \emph{g - r} vs. \emph{i}, \emph{r - i} vs. \emph{i} and
\emph{r - z} vs. \emph{i} using 47 clusters with spectroscopic
redshifts ($0.05 < z < 0.9$) selected from the SPT-SZ survey
\citep{reichardt12,ruel13} and processed in a similar fashion to the BCS
survey data.
For each cluster we first identify an excess of galaxies in
color-magnitude space around the SPT position. We compare the colors
and magnitudes of these galaxies against our red-sequence model as a
function of redshift. To avoid the redshift determination being
influenced by outliers or dominated by a few galaxies with small
photometric uncertainties we bootstrap resample the galaxies and
clip galaxies with colors further than three sigma from the median
offset from the red-sequence color-magnitude relation under
consideration.  For each bootstrap, the estimated redshift is the
redshift at which the $\chi^{2}$ statistic:
\begin{align*}
\chi^{2} = \sum\limits_\mathrm{galaxies} \frac{[Model(magnitude, \
  color, \ z) -
  {\bf g} ]^{2}}{color \ error^{2} + \sigma_\mathrm{rs}\ ^{2}}
\end{align*} 
(where {\bf g} encodes the color and magnitude of the galaxies and
\emph{$\sigma_\mathrm{rs}$} = 0.05 \citep{koester07,mei09} is the intrinsic
spread of the red sequence) is minimized.  We report the redshift as
the median redshift of 100 bootstrap resamples.

\begin{figure}
\epsscale{1.0}
\plotone{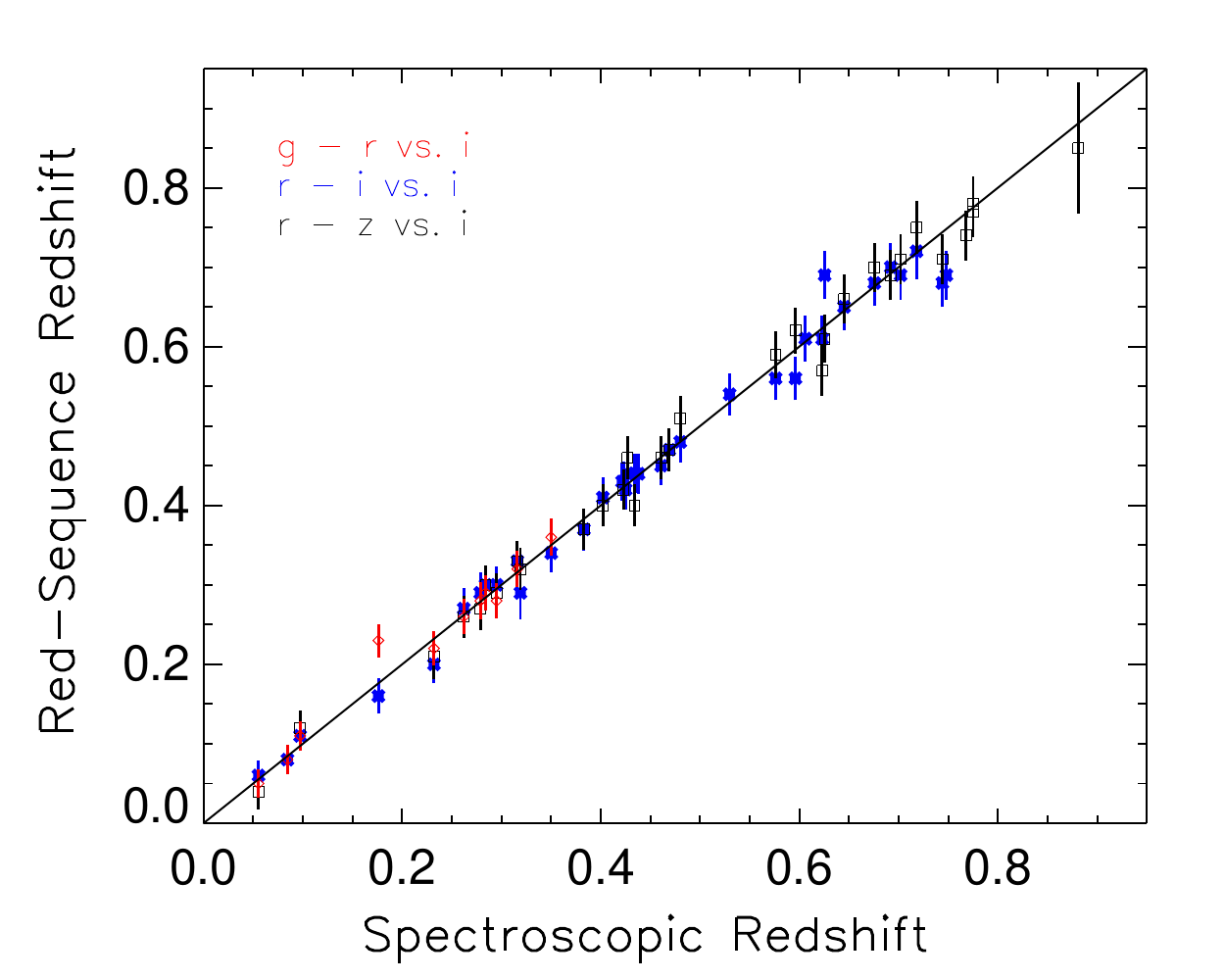}
\caption{Results of red-sequence model training with a sample of
spectroscopic clusters from the SPT-SZ survey. Plotted is the
estimated versus spectroscopic redshift for \emph{g-r vs. i-}band (red
diamonds), \emph{r-i vs. i-}band (blue crosses) and \emph{r-z
vs. i-}band (black boxes). The typical scatter, accounting for the
degrees of freedom removed by the model fitting, is $\delta_z$/(1 + z)
$\sim$ 0.018.}
\label{fig:spectuning}
\end{figure}

We find a simple linear fit of model redshifts $z_{model}$ to
spectroscopic redshifts, $z_{spec}$,
\begin{align*}
  z_\mathrm{spec} =  Az_\mathrm{model} + B 
\end{align*}
is sufficient for tuning the \emph{g - r vs. i} relation over the
redshift range 0.05 to 0.35. However, when extending the redshift
range to 0.75 (\emph{r-i vs. i}) and 0.9 (\emph{r-z vs. i}), structure
is apparent in the residuals from the linear fits. We instead
calibrate these two relations as follows. As we do not, \emph{a
  priori}, have a model for how to map the raw-model redshifts into
the measured spectroscopic redshifts (only the expectation that such a
mapping should be smoothly varying and monotonic),we use non-linear
least squares minimization to fit the $z_{model}$ and $z_{spec}$
relation to a monotonic function generated using the methodology of
\citet{ramsay98} where we have chosen sines and cosines as the basis
functions and include these functions to the 4th order. In Figure
\ref{fig:spectuning} we plot the results of our model calibration.

We estimate the uncertainty in our model calibration by determining
the quantity $\delta z$ such that the reduced chi-squared statistic,
$\chi_\mathrm{red}^{2}$ :

\begin{align*}
 \chi_\mathrm{red}^{2}  = \frac{1}{\nu}  \sum \frac{ (z_\mathrm{estimated} - z_\mathrm{spec})^{2}}{ (\delta z(1 + z))^{2} }  = 1
\end{align*}

where $z_\mathrm{estimated}$ is our calibrated model redshift and $\nu$ is
the number of degrees of freedom. Here the total degrees of freedom
are reduced by 2 by the linear fit in the \emph{g - r vs. i}
calibration and by 10 for the \emph{r -i vs. i} and \emph{r -z
vs. i} calibration. We find uncertainties of $\delta_{z}$/(1 + z)
$\sim$ 0.015-0.018 for all of the redshift models.

\begin{figure*}
 \includegraphics[trim=1.4cm 7cm 1cm 7cm, clip=true, scale=0.75]{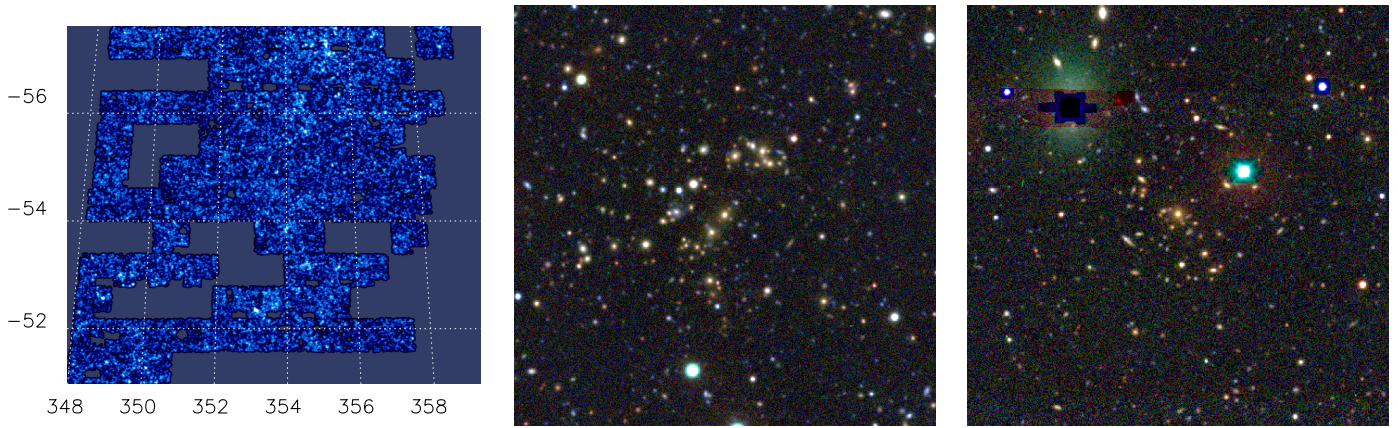}
 \caption{\emph{(Left)} Example red-sequence significance map at
   z=0.47 from the low-z cluster search in the 23 h field. Masked
   stars and amplifiers create the holes in the map.  The color scale
   ranges from -2.5 to 3.3 $\sigma$.  \emph{(Middle)}
   LCS-CLJ231509-5233.6 (SCSO-J231511-523322), a rich cluster from
   this redshift slice at z=0.47 with richness, $\lambda$(0.4\lstar),
   of 27. \emph{(Right)} A more typical cluster, LCS-CLJ233818-5238.1,
   at z=0.54 and richness $\lambda$(0.4\lstar)=10.5.}
\label{fig:cluster_slice}
\end{figure*}

As an illustration of the cluster-detection process, a redshift slice
of the 3d significance data cube from the 23 h field and a pair of
identified clusters is shown in Figure \ref{fig:cluster_slice}.

\section{Cluster Characterization}
\label{sec:remeasure}
Following the initial cluster extraction we determine several
properties for each cluster: its redshift, brightest cluster galaxy
(BCG), and optical richness.  For our richness measure we utilize the
$\lambda$ statistic introduced in \citet{rozo09} and explored further
in \citet{rykoff12}. 

\subsection{Redshift Remeasurement}
We first remeasure the redshift of each cluster.  To determine the
redshift we select all galaxies that contribute to the cluster
detection and then compute the cluster redshift as described in \S
\ref{sec:cluster}. We use the average of \emph{r-i vs i} and  \emph{r-z vs i} red-sequence redshifts
 to determine the redshift for all systems. When this estimated redshift
is $\leq$0.35 we further refine the estimate using the \emph{g-r vs i}
red-sequence model.  The redshift uncertainty is reported as the rms
of 100 bootstrap--estimated redshifts added in quadrature with a
minimum scatter determined during both our spectroscopic tuning and
catalog merging process (see below) of $\delta_z$/(1 + z) = 0.02.  In
general we find good agreement with the initial redshift estimate from
the cluster finder itself except at the boundaries of the searched
redshift-range, or, for systems selected during of the \emph{g-r vs. i} cluster
search, where initial redshifts were misestimated owing to
color-redshift degeneracies.

\subsection{BCG Selection}
Following measurement of the cluster redshift, we next determine the
BCG of each system.  For the BCG we pick the brightest galaxy within a
box in color-magnitude plane such that its \emph{i}-band magnitude,
\emph{m}, is -3 + \mstar$ <$ \emph{m} $<$ \mstar, its \emph{r-z} color
is greater than 2.5 $\sigma$ redward of the red-sequence model and
less than 3.5 $\sigma$ or 0.3 mags (whichever is larger) blueward, and
require the galaxy to be located within 350 kpc of the cluster-finder
determined location. If no such galaxy exists, we expand the search
radius to 450 kpc. If no galaxy is located in this expanded search, we
report the cluster location at the cluster-finder determined location
and flag the system. Otherwise we report the BCG position as the
location of the cluster.
After this automated selection we visually inspect all chosen BCGs and 
manually select an alternative BCG for $\sim$ 30 systems where the BCG was misidentified owing to artifacts, 
deblending issues in the cluster core or lack of inclusion in the catalog owing the the presence of bright stars.

The search radii was chosen based upon a coarse optimization in
simulations as well as visual inspection of the selected
centers in image cutouts around the clusters.  We chose a fixed radius
independent of cluster size as, while 350 kpc is a large fraction of
$r_{500}$ for smaller clusters, the searched area is small enough that
there are few bright galaxies to misidentify as
the BCG.  This optimization may be too simplistic
and further data is required to test the fidelity of our centering
algorithm. 

\subsection{Richness}
\label{sec:lambda_section}
After BCG selection we determine the optical richness of
each cluster.  For our richness measure we use $\lambda$, a
statistic optimized to minimize the scatter in the  Mass-richness
relation \citep{rykoff12}. Much like the red-sequence cluster finder developed in
\citet{gladders00}, the $\lambda$ statistic includes spatial,
magnitude and color weighting to optimize the contrast of cluster
galaxies against the background \citep{rozo09, rykoff12}. 
Simple Monte Carlo tests, as well as tests with mock
catalogs, show that for BCS-depths accurate richnesses can be
determined out to redshift \emph{z}=0.75 counting galaxies to
0.4\lstar \ and to \emph{z}=0.55 counting to 0.2\lstar.  As the
\emph{i-}band data is significantly deeper than the other bands for
red-sequence objects, the relatively larger photometric uncertainties
in the \emph{r-}band determine these limits rather than incompleteness
in the catalogs. 

To compute $\lambda$ we modify the code provided for the SDSS dataset
by
\citet{rykoff12}\footnote{http://kipac.stanford.edu/maxbcg/lambda\_richness.pro}
to utilize our red-sequence model and measurements of the background
source density in color-magnitude space. We estimate two backgrounds
for the BCS, one for each field, as the fields are centered at
different galactic latitudes ($-33$ latitude for the 5 h field and $-58$
latitude for the 23 h field) and our star-galaxy separation does not
extend to the faint limits of the BCS catalogs.  We report $\lambda$
using the \emph{r-i} vs. \emph{i} red-sequence relation as this
color-magnitude combination has the smallest photometric errors in the
BCS. For each cluster, we use the best fit \emph{r-i} vs \emph{i}
red-sequence model, but enforce \lstar \ limits using the best-fit
redshifts as determined above. 

\subsection{Stellar Contamination of the Cluster Catalog}

With the fairly conservative bright star cut of \emph{i-}band =
20.5, the excision of stars from the source catalogs is
not complete to the magnitude limit used for the cluster search. While
the ratio of stars to galaxies falls at fainter magnitudes, the presence of stars (especially class-M stars which
have similar colors and magnitudes as high-redshift cluster galaxies)
can add scatter to richness estimates and, because of the steeply
rising number density of clusters with decreasing richness, can
artificially boost the number of systems above a fixed richness
threshold.  We explore the effect of stellar contamination using the
90\% (96\%) of the tiles searched for clusters (high-redshift
clusters) for which the star separation was robust to \emph{i}=21.5
(sufficient to excise stars to $m_{\star}$+1.0 at $z=0.55$ compared to
$z=0.38$ for the conservative cut). We test for the effect separately
in the 23 h and 5 h fields as we expect the effect to be more
pronounced in the lower galactic latitude 5 h field. 

While some differences in richness may be caused by the removal of actual
cluster galaxies with this stricter cut (as this is at the faint limit
of the star-galaxy separation), our tests in the Extended Groth Strip
show that this leakage should be small.  We can empirically test for
the removal of cluster galaxies at the fainter magnitudes of the cut
using a sample of 11 moderately rich clusters $\lambda(0.4L_{star}) >
20$ at 0.38 $< z <$ 0.43.  At these redshifts the deeper star-galaxy
limiting magnitude reaches $0.2L_{\star}$ at z=0.43 but only
$0.4L_{\star}$ for the shallower cut, so, by comparing the difference
in richness measures between $\lambda(0.2L_{\star})$ and
$\lambda(0.4L_{\star})$ for the two cuts we can estimate the fraction
of cluster galaxies erroneously removed by the deeper cut. The
observed effect is small: from this test we find the mean difference
in richness is -0.17 $\pm$ 1 counts or 0.3\%.

Comparing the numbers of clusters at $\lambda(0.4 L_{\star}) > 10 $ and
$0.55 < z < 0.75$ in the 5 h field we find 272 systems in the deeper
star-cut catalogs opposed to 282 in the shallow. Similarly in the 23 h 
field we find 135/138 clusters in the shallower/deeper-cut
catalogs. While the fixed magnitude cuts affect a declining fraction
of the source population as the redshift
increases, (i.e., with a fixed magnitude limit we can excise stars to
a magnitude comparable to an $m_{\star} + 1$ galaxy at $z=0.55$ but only to
$\sim m_{\star}$ at $z=0.75$), this test shows that \emph{on average} the
background correction is sufficient.  However, to reduce scatter and
mitigate the effects of stellar-contamination we estimate $\lambda$
with deeper stellar-excised catalogs where possible.

After the $\lambda$ computation, we merge the catalogs from the low-
and high-z cluster searches. For every candidate we search for other
candidates within the aperture determined by the $\lambda$ algorithm
(see Equation 4 of \citet{rykoff12}), and within $\Delta$z =0.05 in
redshift. For all matched pairs of candidates we robustly estimate the
scatter in the redshift differences between the pairs, finding a
spread of $\delta_z$/(1 + z) = 0.011 in the difference. As many of
the member galaxies are in common, we consider this extra scatter to
be added by the cluster-finding process and so add it in
quadrature with the model uncertainty determined during the spectroscopic
tuning process. This produces a floor in the redshift uncertainty of
$\delta_z$/(1 + z) = 0.02\footnote{We find similar scatter in the mock
  catalogs (\S \ref{sec:tests}), where the initial tuning with 30 rich
  systems showed a scatter of $\delta_z$/(1 + z) = 0.007 and the final
  catalog showed a net scatter of $\delta_z$/(1 + z) = 0.014 around
  the true redshifts, an extra scatter of 0.012$\times$(1+z) . }.  We
next match clusters within the richness aperture and 3 $\sigma$ of the
redshift scatter from the matched pairs.  As this merger procedure
is designed to both merge separate cluster catalogs and to remove
sub-clumps of rich clusters from the final catalog, we select the
richer of the two systems when duplicates occur. Finally, we make a
final visual inspection of the $\sim$20 clusters with potential
counterparts within 1\arcmin \ but outside the redshift cut.  We
exclude the few systems which are clearly composed of a small subset
of galaxies from a richer lower-redshift system or clearly mismeasured
at the boundary redshift of the low-z search.

\subsection{Compensating for Masked Regions}

As a last step, we compensate for the systematic reduction in richness
owing to masked regions around the clusters. For every cluster we
compute the fraction of area masked around the BCG in 100 kpc-wide
rings out to 2 Mpc. Using the galaxy weights provided by the
$\lambda$-algorithm we then compute an area-corrected
richness-per-radial bin for each cluster for both
$\lambda(0.4L_{\star})$- and
$\lambda(0.2L_{\star})$-richnesses. Combining the cluster catalog from
both fields to improve statistics and excluding the most heavily
masked systems, we use clusters between 0.3 $< z < 0.75$ (0.55 for
$\lambda(0.2L_{\star}$)) to compute an ``average'' richness-per-radial
bin as a function of richness.  This average is derived from hundreds
of systems at the low-richness end down to $\sim$10 clusters for the
richest systems. Based upon an individual cluster's masking we then
compute a correction to its richness. If this richness correction
pushes the system into a higher-richness bin, we iterate the
correction process until the correction converges. We report the
corrected richnesses as well as the value of the  richness-correction as entries in
the cluster tables.

\section{Tests on Simulated Catalogs}
\label{sec:tests}

We characterize the completeness and purity of the cluster-detection
algorithm using simulated galaxy catalogs.  As we have not
incorporated the complex masking required for the BCS into these tests
we consider these results to be an idealized scenario.

The mock galaxy sample is drawn from a 220 degree lightcone populated with galaxies in the redshift range 0 $< z <$ 1.3 down to a flux limit $m_i \sim 25$.  
The underlying dark matter distribution is based on a cosmological simulation of 1 $h^{-1}$Gpc; this simulation is a single `Carmen' simulation from the Large Suite of Dark Matter Simulations project \citep[LASDAMAS][]{Mcbride++11}\footnote{http://lss.phy.vanderbilt.edu/lasdamas}.  
The lightcone was created by pasting together 34 separate snapshots.
The Adding Density Determined GAlaxies to Lightcone Simulations (ADDGALS) algorithm is run to assign galaxies to the dark matter particles in a way that reproduces the known luminosities and two-point function.  
SEDs are then added using a training set of low-redshift spectroscopic galaxies from SDSS DR5 in such a way as to reproduce the observed magnitude-color-environment relation. 
These simulated catalogs produce realistic distributions of galaxies and their colors, including a well defined cluster red sequence. 
Because of these properties, the catalogs have previously been used for tests of cluster finding \citep{koester07, hao10, Soares-Santos++11, Rykoff++13}, photometric redshifts \citep{gerdes10}, and spectroscopic followup \citep{Cunha++12}. 
The technique was previously presented in  \citet{Wechsler04, BushaWechsler08}, and a full description of the algorithm and the simulated sky catalogs it produces will be presented in Wechsler et al.~(in preparation) and Busha et al.~(in preparation).

We degrade the mocks to BCS depths by adding to the
model fluxes a random deviate drawn from the normal
distribution, N(0,$\sigma_\mathrm{BCS}$),  where $\sigma_\mathrm{BCS}$ is the typical source flux uncertainty in each
optical band: 0.19, 0.21, 0.35, and 1.3 uJy in the
\emph{g-,r-,i-,} and \emph{z-}bands respectively: 

\begin{align}
 flux =  flux_\mathrm{true} +  N(0, \sigma_\mathrm{BCS}).
\end{align}

The fluxes are converted to magnitudes and sources with \emph{i-}band
magnitude uncertainty $>0.3$ or \emph{r-, z-}band uncertainties $>0.5$
are excluded.  In this simplified degradation of the mocks, we do not
attempt to include the effects of incompleteness beyond these
uncertainty constraints.

We tune our synthetic red-sequence model to the simulations in an
analogous procedure to that used to connect these models to the real
cluster observations. We select 30 halos with redshift, $z_{halo}$,
$0.09 > z_{halo} > 0.9$ and \mtwohundred $>3\times 10^{14} \msun$ from the
mock catalogs.

As in \S \ref{sec:catalog} we identify an excess of red-sequence
galaxies around these systems and use a monotonic remapping of our
synthetic redshift $z_{model}$ to the halo redshifts.  Using these red
sequence models we run the cluster-finding algorithm on the
simulations, where we have split the simulated catalog into ten
patches of 22 \degs \ (or roughly the scale of each patch used in the
BCS cluster-search).

\begin{figure*}
\epsscale{1.0}
\includegraphics[width=\textwidth]{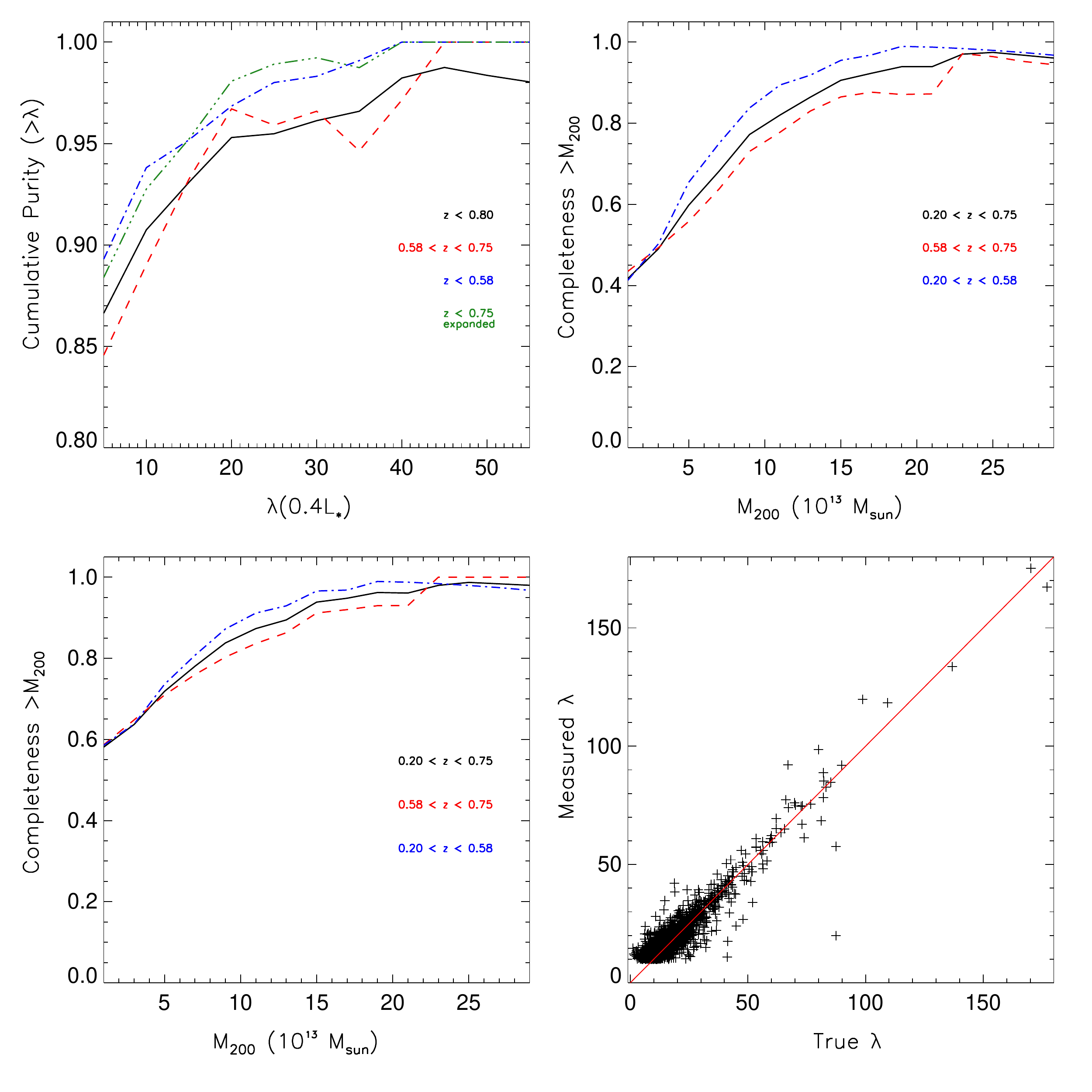}
\caption{Purity and completeness tests of the cluster-detection
  algorithm on mock catalogs (\S \ref{sec:tests}). (\emph{Top Left})
  Fraction of systems matched to halos within $r_{200}$ and $\Delta$z
  $< 0.035\times (1+z_{halo})$ for 3 redshift ranges: z$<0.8$
  (solid-black), $0.58<z<0.75$ (dashed-red), $z<0.58$ (dot-dashed
  blue) where 0.58 is the median redshift of the extracted cluster
  catalog. Expanding the redshift matching criterion to $\Delta$ z $<
  0.05\times (1+z_{halo})$ produces the green dashed-triple dotted
  line. (\emph{Top Right}) Fraction of halos recovered with $>10$
  members (both red and blue) where matched clusters are centered
  within $r_{200}$ of the halo and $\Delta$z $< 0.035\times
  (1+z_{halo})$. (\emph{Bottom Left}) Identical to top right panel except
  plotting fraction of recovered halos with $\lambda_{true} >10$ where
  $\lambda_{true}$ is measured to 0.4L$_{\star}$ centered on the halo
  center and using the true magnitudes for both the lambda measurement
  and background estimation.  (\emph{Bottom Right}) $\lambda_{true} $
  versus measured $\lambda$ for matched systems at $z < 0.75$ for
  halos with $ \mtwohundred > 2 \times 10^{13} \msun$.}
\label{fig:simulation_check}
\end{figure*}

We make the same cuts on the recovered cluster catalog as were used in
constructing the BCS cluster catalogs --- namely signal-to-noise
$>3.1$ and $\lambda(0.4L_{\star}) > 10$. To check the purity of this
catalog we cross-match against the simulation halo catalog.  Here we
adopt a similar matching criterion to \citet{dong07} and consider a
cluster to match a halo if the projected distance between the cluster
and halo center is less than $r_{200}$ and the redshift difference,
$\Delta$z, is $< 0.035\times (1+z_{halo})$.  The results of these
checks are shown in Figure \ref{fig:simulation_check}.  We find an
accuracy in the recovered redshifts on the simulations to have scatter
of $\delta_{z}$/(1 + z) = 0.014.

We find $>85$\% purity for systems with $\lambda >10 $ out to redshift
0.75.  Expanding the redshift scaling to $0.05\times (1+z_{halo})$ to
account for outliers, the purity exceeds 90\%. Beyond z=0.75 the
purity falls off, primarily due to increasing discrepancies between
the model redshift and the simulation red-sequence.  As the
photometric uncertainties in the BCS preclude accurate richness
measurements above z=0.75, we reserve exploration of the
cluster-finder at higher redshifts for future work. 

Next, considering halos with more than 10 members of any magnitude and
color (ie both blue and red members), the catalog is 90\% complete
above 2$\times 10^{14} \msun $.  Finally, cutting the
halo catalogs at $\lambda_{true} > 10$ (where $\lambda_{true}$ is
measured for all halos to 0.4$L_{\star}$ by centering on the halo's
central galaxy and using the true galaxy magnitudes and redshifts) the
cluster catalogs are 90\% complete at 1$\times 10^{14} \msun $. Exploring the missing halos above this mass we find
75\% of the missing systems exceed our cluster extraction threshold in
the cluster-finder datacube. As such, these systems are either falling
outside our redshift search, below the richness threshold or being
merged with a nearby system.

\section{Final Catalog}
\label{sec:compare}

\begin{figure}
\epsscale{0.75}
\plotone{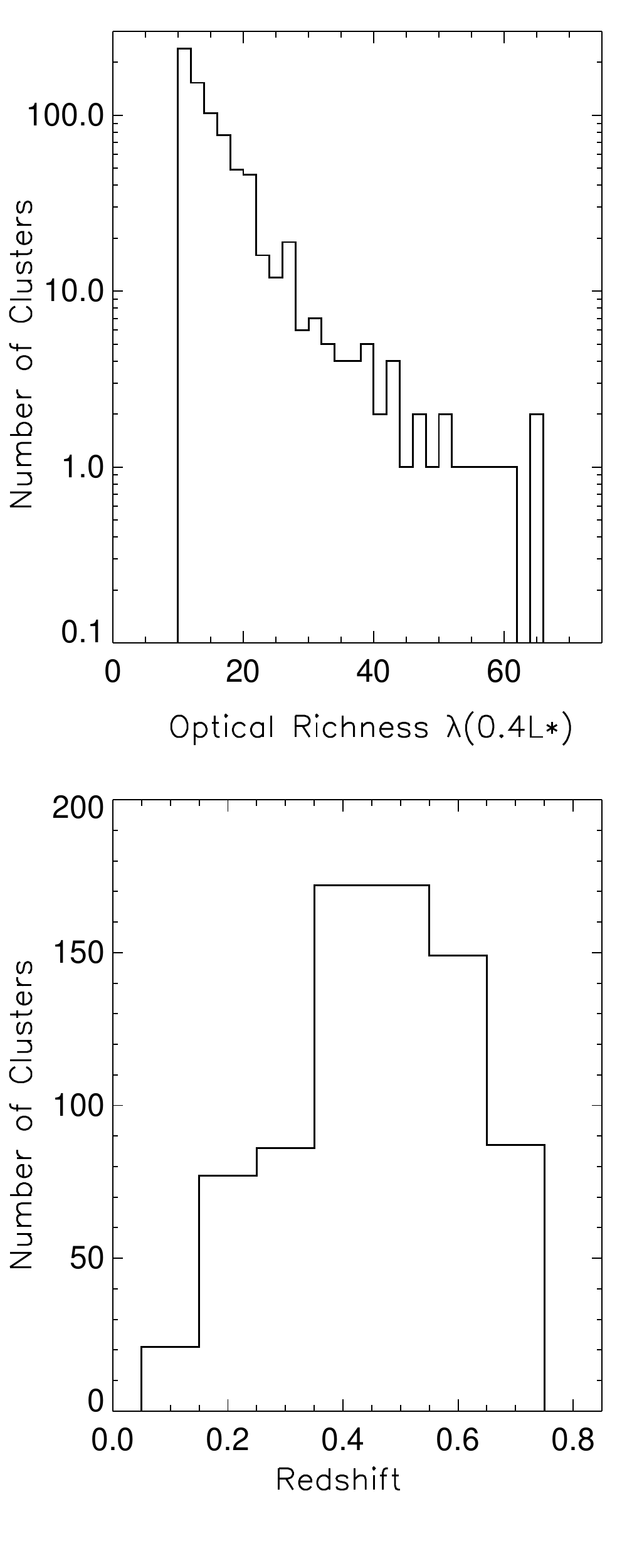}
\caption{\emph{(Top)} Distribution in optical richness, $\lambda$, measured to
  $0.4L_{\star}$ for the \numberclusters \ clusters at z $\leq$
  0.75. \emph{(Bottom)} Redshift distribution of the entire cluster
  catalog. }
\label{fig:cluster_stats}
\end{figure}

The final cluster catalog extracted from the BCS consists of
\numberclusters \ clusters (\numberclustersfive \ and
\numberclusterstwenty \ clusters in the 5 and 23 h fields,
respectively) at $\lambda \geq 10 $ and at z $\leq$ 0.75. The median
redshift of the sample is \medianredshift \ and the median richness is
\medianrichness.  More than 85\% of the sample is newly
discovered. We plot the redshift and richness distribution of
this catalog in Figure \ref{fig:cluster_stats}.  We denote these systems by
the acronym ``LCS'', for Little Cluster Survey, in light of the
ongoing Dark Energy Survey\footnote{www.darkenergysurvey.org} which is
in the midst of a 5 year survey that will image 5000 \degs \ of the
southern sky (including the BCS region). In Tables
\ref{tab:clusters}--\ref{tab:clusters2} we report the locations,
redshifts, and richnesses of the systems as well as previous
identifications in the literature.

\subsection{Comparisons to Other Cluster Catalogs}
Here we compare our cluster catalog with other catalogs of galaxy
clusters which contain systems within the BCS footprint.  In this
section we specifically consider systems from the SPT and ACT mm-wave
surveys \citep{vanderlinde10,reichardt12,marriage11b}, the optical Southern
Cosmology Survey (SCS) \citep{menanteau09,menanteau10}, and the X-ray
XMM-BCS survey \citep{suhada12}. (For a more thorough cross-check in
tables \ref{tab:clusters}--- \ref{tab:clusters2}, we also query the
SIMBAD\footnote{http://simbad.u-strasbg.fr/simbad} database and check
for counterparts within 3\arcmin\ of each system). The SZ-selected
clusters are massive, with SZ-determined masses $\mfivehundred >2.8 \times 10^{14}
\msun$ \citep{reichardt12,marriage11b}, 
 the SCS-systems are selected
based upon their galaxy content to have masses $\mtwohundred > 3 \times 10^{14}
\msun $ (with respect to the mean density of the
universe) \citep{menanteau10}, and --as is typical for a flux-limited
X-ray survey-- the XMM-BCS systems span a broad range in mass, with a
median mass of $ \mfivehundred \approx 1 \times 10^{14} \msun$ as derived from the X-ray
luminosities \citep{suhada12}.

For this comparison we only consider systems within the redshift range
searched for a given BCS tile---e.g., we do not consider a
high-redshift system to be ``missed'' in our search if the
cluster-finding algorithm was only run at significantly
lower-redshifts. Taking into account possible differences in
redshift-estimation accuracy we use cuts of $0.15 <z<0.55$ for tiles
used in the low-redshift cluster search and extend this to $z<0.9$ for
regions included in the higher-redshift search. We consider systems to
be a ``match'' if they are within $|\delta z | <$ 0.2 and within the
richness algorithm cutoff radius (typically 900 kpc). We adopt the
large redshift matching-criterion to avoid missing systems due to
differences in estimated redshifts (in particular, for the SCS systems
we find discrepancies between our redshifts and the redshifts reported
in \citet{menanteau10} similar to those reported in \citet{suhada12}).
As expanded upon below, the results proceed as expected: we recover
all non-masked, massive systems from the SZ-surveys, the majority of
the optically-selected clusters from SCS catalog, and a significant
fraction of the XMM-BCS catalog with the non-matched systems generally
residing at lower masses.

The SCS catalog provides the largest catalog of clusters for
comparison --- 65 systems lie within the redshift range and footprint
searched.  Of these 65 clusters, 54 have counterparts in our
catalogs. Of the remaining 11 systems, 4 are matched by clusters
detected by our cluster-finder but at richnesses below our cutoff
threshold, 2 were significantly masked within 1\arcmin \ around
the SCS-BCG by our automated masking procedure, 
and for 2 clusters both algorithms identified the same cluster but selected significantly separated BCGs. 
One cluster was not detected in our search and the two remaining 
SCS-detected clusters were not found as their redshifts (as measured by our algorithms) lie outside of the range searched. 
Inverting the question, for tiles the catalogs share
in common and using $\lambda(0.4L_{\star})$ as the richness metric, we
find SCS counterparts for 20/25 (32/50) of the richest systems in
these regions.

From the XMM-BCS catalog there are 40 systems in this comparison (we
include the ``lower-quality'' detections in this check).  Of these, 23
systems have counterparts in our catalog while an additional 7 were
detected by the cluster-finder but have richnesses below the cutoff
threshold.  For the matched systems we find the redshifts to be on
average consistent within the quoted uncertainties.  The outstanding
10 non-recovered systems (ID 69,70,81,94,156,355,357,536,540) have estimated
masses ranging from $\mtwohundred \sim 4-25 \times 10^{13}
\msun$.  Three of
these systems (355, 536, 540) are termed ``lower-quality'' detections in the X-ray
data \citep{suhada12} and have masses $\mtwohundred < 10 \times 10^{13} \msun$ \ 
and five of these systems (94,156,357,430,540) lie on
the edges of BCS tiles where coverage is less uniform.

Of the 10 possible mm-wave selected-systems (one of which is also a
REFLEX cluster \citep{bohringer04}), 8 have counterparts in our
catalogs while the remaining 2 unmatched systems have greater than
50\% of the area within 1\arcmin \ of the reported center masked owing
to the presence of bright stars. Not surprisingly (as some of these
systems were used in our red-sequence model training), there is good agreement
between our estimated redshifts and the literature redshifts for these
systems.

\section{Summary and Future Work}
\label{sec:discussion}

In this paper we have presented our reductions of optical imaging data
from the Blanco Cosmology Survey, including our methodology for
creating calibrated source catalogs, for correcting underestimated
photometric errors returned by SExtractor, and our implementation of a
new, easily-coded morphological star-galaxy separation statistic. 
Using a red-sequence-based cluster finding algorithm, we have searched
the BCS for galaxy clusters.  We report the coordinates, redshifts,
and optical richnesses for 
\numberclusters \ clusters at z $\leq$\redshiftlimit, of which greater than
\newpercent \% are new detections. This sample has a median redshift of
\medianredshift \ and median optical richness, $\lambda$, of
\medianrichness. Based upon tests with realistic mock catalogs, the
catalog is expected to be $>85\%$ pure at z $< 0.75$. 

One of the strengths of the Blanco Cosmology Survey is its overlap
with other multi-wavelength data. The optical catalogs presented here
have been used to confirm cluster candidates and estimate redshifts
for SZ-selected clusters in the SPT-SZ survey \citep{reichardt12,
  song12b}, as well as to make the first measurement of galaxy bias
from the gravitational lensing of the Cosmic Microwave Background
\citep{bleem12b}. As these reductions might be of broader utility, we release the
reduced \emph{g-}, \emph{r-},  \emph{i-}, \emph{z-}band images and weight maps as well as the
calibrated source catalogs for this \bcssize \ survey.  These products
are available at  \webaddress.

\begin{acknowledgements}
  The authors wish to thank Michael Huff for assistance with flagging
  spurious objects in the source catalogs and Doug Rudd for his assistance setting up the online data access. 
  LB would like to thank Tom Crawford for useful discussions.  
  LB acknowledges support by the U.S. Department of Energy, Basic Energy Sciences, Office of Science, under contract No. DE-AC02-06CH11357, the
  NSF Physics Frontier Center award PHY-0551142 and the NSF OPP award
  ANT-0638937.  Galaxy cluster research at SAO is supported in
  part by NSF grants AST-1009649 and MRI-0723073.
  This research draws upon data provided by NOAO PI 2005B-0043 as
  distributed by the NOAO Science Archive. NOAO is operated by the
  Association of Universities for Research in Astronomy (AURA),
  Inc. under a cooperative agreement with the National Science
  Foundation. Characterization of the new star-galaxy classifier was
  based upon both data from both AEGIS (a multi-wavelength sky survey
  conducted with the Chandra, GALEX, Hubble, Keck, CFHT, MMT, Subaru,
  Palomar, Spitzer, VLA, and other telescopes and supported in part by
  the NSF, NASA, and the STFC) and upon observations obtained with
  MegaPrime/MegaCam, a joint project of CFHT and CEA/DAPNIA, at the
  Canada-France-Hawaii Telescope (CFHT) which is operated by the
  National Research Council (NRC) of Canada, the Institut National des
  Science de l'Univers of the Centre National de la Recherche
  Scientifique (CNRS) of France, and the University of Hawaii. This
  work is based in part on data products produced at the Canadian
  Astronomy Data Centre as part of the Canada-France-Hawaii Telescope
  Legacy Survey, a collaborative project of NRC and CNRS.
  Additionally, this research has made use of the SIMBAD database,
  operated at CDS, Strasbourg, France. 
  Finally, the authors acknowledge the University of Chicago Research Computing Center for hosting the data products presented in this work.

\end{acknowledgements}

{\it Facilities:}
\facility{Blanco (MOSAIC II)}

\bibliography{bcs}

\begin{appendix}

\section{BCS Tile Information}

Summary information for reduced BCS tiles. Images and calibrated
catalogs are available at \webaddress.

\LongTables
\pagestyle{empty}
\def\arraystretch{1.2}
\tabletypesize{\footnotesize}
\begin{deluxetable}{ lcc | cccc | cccc  | c  | l }
\tablecaption{Summary information for BCS Data presented in this work.
\label{tab:tile}
}
\tablehead{
\multicolumn{3}{c}{\bf{ID \& coordinates:}}  & 
\multicolumn{4}{c}{\bf{Seeing (arcsec)}}  &
\multicolumn{4}{c}{\bf{5$\sigma$ Point Source Depth}}  &
\multicolumn{1}{c}{\bf {Star Limit}}  &
\multicolumn{1}{c}{\bf {Comments}} \\
\colhead{BCS ID}  & 
\colhead{RA}  & 
\colhead{DEC}   & 
\colhead{\emph{g}}  & 
\colhead{\emph{r}}  & 
\colhead{\emph{i}}  & 
\colhead{\emph{z}}  & 
\colhead{\emph{g}}  & 
\colhead{\emph{r}}  & 
\colhead{\emph{i}}  & 
\colhead{\emph{z}} 
}

\startdata 
BCS0506-5601 & $ 76.708 $ & $ -56.015 $ & $ 1.33 $ & $ 1.23 $ & $ 1.11 $ & $ 0.99 $ & $ 24.19 $ & $ 23.88 $ & $ 23.35 $ & $ 21.65 $ & $ 21.78$ & - \\
BCS0507-5412 & $ 76.985 $ & $ -54.199 $ & $ 0.91 $ & $ 1.01 $ & $ 0.75 $ & $ 0.68 $ & $ 24.35 $ & $ 24.47 $ & $ 23.89 $ & $ 22.34 $ & $ 22.21$ & L,H \\
BCS0507-5448 & $ 76.897 $ & $ -54.803 $ & $ 0.95 $ & $ 0.92 $ & $ 0.78 $ & $ 0.76 $ & $ 24.38 $ & $ 24.30 $ & $ 23.80 $ & $ 22.26 $ & $ 22.23$ & L,H \\
BCS0507-5524 & $ 76.804 $ & $ -55.411 $ & $ 1.07 $ & $ 1.10 $ & $ 0.90 $ & $ 0.96 $ & $ 24.28 $ & $ 24.24 $ & $ 23.60 $ & $ 22.05 $ & $ 22.24$ & L,H \\
BCS0508-5223 & $ 77.195 $ & $ -52.398 $ & $ 1.10 $ & $ 0.93 $ & $ 0.91 $ & $ 0.89 $ & $ 24.17 $ & $ 24.10 $ & $ 23.97 $ & $ 22.37 $ & $ 22.09$ & L,H \\
BCS0508-5300 & $ 77.174 $ & $ -52.995 $ & $ 1.12 $ & $ 1.10 $ & $ 0.88 $ & $ 0.79 $ & $ 23.89 $ & $ 24.15 $ & $ 23.78 $ & $ 22.20 $ & $ 22.02$ & L,H \\
BCS0508-5336 & $ 77.067 $ & $ -53.597 $ & $ 0.88 $ & $ 0.98 $ & $ 0.91 $ & $ 0.82 $ & $ 24.15 $ & $ 24.19 $ & $ 23.69 $ & $ 22.13 $ & $ 21.98$ & L,H \\
BCS0509-4959 & $ 77.482 $ & $ -49.977 $ & $ 1.46 $ & $ 1.49 $ & $ 0.83 $ & $ 0.76 $ & $ 24.27 $ & $ 23.95 $ & $ 23.94 $ & $ 22.32 $ & $ 21.71$ & L,H \\
\enddata
\tablecomments{Summary information for  the BCS imaging data presented
  in this work.  Coordinate positions are given for  the center of the
  36\arcmin \  $\times$ 36\arcmin \  tiles. The  reported seeing  is the
  average full width at half maximum  of the seeing disk of the single
  epoch images that contribute to the coadded images. The point source
  depths are calculated as in Section \S\ref{sec:catalog} and the Star
  Limit corresponds to the  faintest \emph{i}-band magnitude for which
  we  find robust morphological  separation of  stars and  galaxies as
  described  in Section  \S\ref{sec:stargal}. In  the ``Comments''
  field , ``L''  corresponds  to  tiles utilized  in  the  low-redshift
  cluster   search   while    ``H''   denotes   tiles   searched   for
  higher-redshift systems (see \S \ref{sec:cluster}).}
\end{deluxetable}
\clearpage

\begin{deluxetable}{lcl}
\tablecolumns{3}
\tablecaption{Description of Catalog Columns}
\label{tab:source_catalog_inputs}
\startdata 
   FIELDNAME & - & BCS tile name      \\
   OBJ\_ID   &  - &  Object Identification Number (unique per tile) \\
   RA       & degree & Right Ascension (J2000)       \\
   DEC      & degree & Declination (J2000) \\
   G        & AB Magnitude & \emph{g}-band MAG\_AUTO        \\
   GERR     & AB Magnitude & \emph{g}-band corrected magnitude uncertainty  (See \S \ref{sec:catalog})     \\
   G\_SEX    & - & \emph{g}-band Source Extractor Flags + Bright Star proximity flag\tablenotemark{(a)}     \\
   G\_CLASSSTAR     & - &  \emph{g}-band Source Extractor Class Star  \\
\enddata

\tablecomments{
 $^{\rm (a)}$ Reported flags are the sum of all possible extraction flags (as is standard in SExtractor output).  \\
  0 --- 124  Standard SExtractor flags \\
  256   Missing data \\
  512   Corrupted \emph{i}-band (i.e. bright source that saturated in \emph{i}-band, but not \emph{r-} or \emph{z-}band) \\
  1024  Photometry potentially corrupted owing to proximity to bright star \\
}

\end{deluxetable}

\end{appendix}

\clearpage

\LongTables
\pagestyle{empty}
\begin{landscape}
\begin{center}
\def\arraystretch{1.2}
\tabletypesize{\scriptsize}

\begin{deluxetable}{lllccccccccccccc}  
\tablecolumns{16}
\tablecaption{Optically-Selected Galaxy Clusters located in the 23 h BCS field
\label{tab:clusters}
} \tablehead{ 
  \colhead{ID} & 
  \colhead{R.A.} &
  \colhead{Dec} &
  \colhead{z} &
  \colhead{$\delta$z} &
  \colhead{Sigma} & 
  \colhead{$\lambda$} &
  \colhead{$\delta \lambda$} & 
  \colhead{$\Delta \lambda$} &
  \colhead{$\lambda(0.2 L_{\star})$} & 
  \colhead{$\delta \lambda (0.2 L_{\star})$} &
  \colhead{$\Delta \lambda(0.2 L_{\star})$} &
  \colhead{Mask 200 kpc}  &
  \colhead{Mask 500 kpc}  &
  \colhead{Edge of Tile} &
  \colhead{Previous ID} } 
\startdata
LCS-CL J231218-5109.2 & $348.0766 $ & $ -51.1548 $ & $ 0.32 $ & $ 0.025 $ & $ 3.21 $ & $ 14.4 $ & $ 2 $ & $ 1.3 $ & $ 23.8 $ & $ 3 $ & $ 2.6 $ & $ 0.00 $ & $ 0.05 $ & $ 1$ & - \\
LCS-CL J231224-5115.3 & $348.1009 $ & $ -51.2560 $ & $ 0.75 $ & $ 0.040 $ & $ 3.27 $ & $ 14.0 $ & $ 3 $ & $ 0.0 $ & $ - $ & $ - $ & $ - $ & $ 0.00 $ & $ 0.01 $ & $ -$ & - \\
LCS-CL J231230-5214.0 & $348.1262 $ & $ -52.2335 $ & $ 0.28 $ & $ 0.024 $ & $ 4.14 $ & $ 14.1 $ & $ 2 $ & $ 1.3 $ & $ 18.3 $ & $ 3 $ & $ 1.9 $ & $ 0.00 $ & $ 0.03 $ & $ 1$ & - \\
LCS-CL J231235-5218.6 & $348.1469 $ & $ -52.3114 $ & $ 0.42 $ & $ 0.031 $ & $ 4.11 $ & $ 16.3 $ & $ 2 $ & $ 0.9 $ & $ 34.0 $ & $ 4 $ & $ 2.2 $ & $ 0.00 $ & $ 0.03 $ & $ -$ & - \\
LCS-CL J231240-5201.3 & $348.1674 $ & $ -52.0220 $ & $ 0.49 $ & $ 0.033 $ & $ 3.65 $ & $ 11.7 $ & $ 2 $ & $ 0.4 $ & $ 17.2 $ & $ 3 $ & $ 0.6 $ & $ 0.00 $ & $ 0.04 $ & $ -$ & - \\
LCS-CL J231253-5203.4 & $348.2222 $ & $ -52.0570 $ & $ 0.69 $ & $ 0.040 $ & $ 3.51 $ & $ 12.1 $ & $ 2 $ & $ 0.1 $ & $ - $ & $ - $ & $ - $ & $ 0.00 $ & $ 0.00 $ & $ -$ & - \\
LCS-CL J231305-5259.0 & $348.2729 $ & $ -52.9843 $ & $ 0.65 $ & $ 0.037 $ & $ 3.93 $ & $ 11.0 $ & $ 2 $ & $ 0.5 $ & $ - $ & $ - $ & $ - $ & $ 0.01 $ & $ 0.01 $ & $ -$ & - \\
LCS-CL J231309-5101.5 & $348.2886 $ & $ -51.0254 $ & $ 0.33 $ & $ 0.024 $ & $ 4.67 $ & $ 21.0 $ & $ 2 $ & $ 1.1 $ & $ 37.1 $ & $ 4 $ & $ 2.7 $ & $ 0.04 $ & $ 0.03 $ & $ -$ & - \\
\enddata 

\tablecomments{Galaxy Clusters selected above 3.1 $\sigma$ with
  $\lambda(0.4L_{\star})$ greater than 10 and z $ \le 0.75$ in
  the 23 h field.  We report the cluster
  ID, the BCG position, red-sequence redshift and uncertainty and
  detection significance. For richnesses derived to 0.4$L_{\star}$ and
  0.2$L_{\star}$ (for z $< 0.55$) we include the optical richness
  ($\lambda$), richness uncertainty and the amount the richness was
  boosted to account for masking. Additionally, we report the fraction
  of area masked within 200 and 500 kpc, flag the system if it is
  located on the edge of a tile, and note if the cluster has a
  previous identification.  The reported richness uncertainty is the
 uncertainty  estimate from the richness algorithm (see Equation 3 of
  \citet{rykoff12})  added in quadrature with a Poisson uncertainty from masked-area compensation. The previous
  identification column designates a counterpart using the criterion
  of \S\ref{sec:compare}. Tags correspond to the following works: \\
  0 ACO \citep{abell89} \\
  1 REFLEX \citep{bohringer04}, \\
  2 Southern Cosmology Survey \citep{menanteau09,menanteau10}, \\
  3 SPT-SZ Survey \citep{vanderlinde10,reichardt12}, \\
  4 ACT \citep{marriage11b}, \\
  5 XMM-BCS \citep{suhada12}, \\
  6 \emph{PLANCK} \citep{planck13-29}  \\
  7 \citet{vikhlinin98} \\
  8 \citet{buckleygeer11}.  }
\end{deluxetable}
\end{center}
\clearpage

\clearpage
\end{landscape}

\begin{landscape}
\begin{center}

\def\arraystretch{1.2}
\begin{deluxetable}{lllccccccccccccc}  
\tabletypesize{\scriptsize}
\tablecolumns{16}
\tablecaption{Optically-Selected Galaxy Clusters located in the 5 h BCS field
\label{tab:clusters2}
} \tablehead{ 
  \colhead{ID} & 
  \colhead{R.A.} &
  \colhead{Dec} &
  \colhead{z} &
  \colhead{$\delta$z} &
  \colhead{Sigma} & 
  \colhead{$\lambda$} &
  \colhead{$\delta \lambda$} & 
  \colhead{$\Delta \lambda$} &
  \colhead{$\lambda(0.2 L_{\star})$} & 
  \colhead{$\delta \lambda (0.2 L_{\star})$} &
  \colhead{$\Delta \lambda(0.2 L_{\star})$} &
  \colhead{Mask 200 kpc}  &
  \colhead{Mask 500 kpc}  &
  \colhead{Edge of Tile} &
  \colhead{Previous ID} } 
\startdata
LCS-CL J050624-5423.5 & $76.6001 $ & $ -54.3925 $ & $ 0.50 $ & $ 0.033 $ & $ 4.56 $ & $ 11.8 $ & $ 2 $ & $ 0.5 $ & $ 12.9 $ & $ 2 $ & $ 0.5 $ & $ 0.00 $ & $ 0.03 $ & $ -$ & - \\
LCS-CL J050715-5145.6 & $76.8148 $ & $ -51.7607 $ & $ 0.64 $ & $ 0.037 $ & $ 3.63 $ & $ 14.5 $ & $ 3 $ & $ 2.4 $ & $ - $ & $ - $ & $ - $ & $ 0.00 $ & $ 0.10 $ & $ 1$ & - \\
LCS-CL J050728-5431.4 & $76.8682 $ & $ -54.5246 $ & $ 0.45 $ & $ 0.033 $ & $ 3.64 $ & $ 11.1 $ & $ 2 $ & $ 1.9 $ & $ 17.8 $ & $ 3 $ & $ 3.0 $ & $ 0.31 $ & $ 0.17 $ & $ 1$ & - \\
LCS-CL J050735-5218.0 & $76.8994 $ & $ -52.3001 $ & $ 0.43 $ & $ 0.031 $ & $ 4.84 $ & $ 11.2 $ & $ 2 $ & $ 0.5 $ & $ 19.5 $ & $ 3 $ & $ 1.0 $ & $ 0.00 $ & $ 0.05 $ & $ -$ & - \\
LCS-CL J050736-5119.8 & $76.9020 $ & $ -51.3314 $ & $ 0.62 $ & $ 0.040 $ & $ 3.39 $ & $ 17.8 $ & $ 3 $ & $ 2.1 $ & $ - $ & $ - $ & $ - $ & $ 0.00 $ & $ 0.01 $ & $ 1$ & - \\
LCS-CL J050805-5217.6 & $77.0214 $ & $ -52.2947 $ & $ 0.61 $ & $ 0.036 $ & $ 3.66 $ & $ 12.1 $ & $ 2 $ & $ 0.7 $ & $ - $ & $ - $ & $ - $ & $ 0.00 $ & $ 0.06 $ & $ -$ & - \\
LCS-CL J050806-5248.4 & $77.0285 $ & $ -52.8081 $ & $ 0.69 $ & $ 0.039 $ & $ 3.66 $ & $ 23.8 $ & $ 3 $ & $ 0.6 $ & $ - $ & $ - $ & $ - $ & $ 0.02 $ & $ 0.00 $ & $ -$ & - \\
LCS-CL J050813-5213.4 & $77.0571 $ & $ -52.2241 $ & $ 0.39 $ & $ 0.035 $ & $ 3.58 $ & $ 11.8 $ & $ 2 $ & $ 1.6 $ & $ 11.5 $ & $ 2 $ & $ 1.5 $ & $ 0.17 $ & $ 0.09 $ & $ -$ & - \\
\enddata 

\tablecomments{The same as Table \ref{tab:clusters}, except for clusters selected in the 5 h field. }
\end{deluxetable}
\end{center}
\clearpage

\end{landscape}

\pagestyle{plain}

\end{document}